\documentclass{acm_proc_article-sp}
\usepackage{amsmath}

\usepackage{stmaryrd}
\usepackage{mathbbold}
\usepackage{amsfonts}
\usepackage{txfonts}
\usepackage{amssymb}
\usepackage{balance}
\usepackage{times}
\usepackage{helvet}
\usepackage{courier}
\usepackage{graphicx}
\usepackage{subfigure}
\usepackage{multirow}
\usepackage{algorithmic}
\usepackage{algorithm}
\usepackage{mathcomp}

\newtheorem{problem}{Problem}
\newtheorem{lemma}{Lemma}
\newcommand{\mat}[1]{{\bf #1}}   
\newcommand{\reminder}[1]{{\bf *** #1 ***}}
\newcommand{\hide}[1]{}
\newcommand{\QED}{\hfill $\Box$ \hfill}

\begin{document}
%
\title{MaTrust: An Effective Multi-Aspect Trust Inference Model}

\numberofauthors{5}

\author{
\alignauthor
Yuan Yao\\
       \affaddr{State Key Laboratory for Novel Software Technology, China}\\
       \email{yyao@smail.nju.edu.cn}
\alignauthor
Hanghang Tong\\
       \affaddr{IBM T.J. Watson Research}\\
       \email{htong@us.ibm.com}
\alignauthor
Xifeng Yan\\
       \affaddr{University of California at Santa Barbara, USA}\\
       \email{xyan@cs.ucsb.edu}
\and
\alignauthor
Feng Xu\\
       \affaddr{State Key Laboratory for Novel Software Technology, China}\\
       \email{xf@nju.edu.cn}
\alignauthor Jian Lu\\
       \affaddr{State Key Laboratory for Novel Software Technology, China}\\
       \email{lj@nju.edu.cn}
}
\maketitle

\begin{abstract}

Trust is a fundamental concept in many real-world applications such as e-commerce and peer-to-peer networks. In these applications, users can generate local opinions about the counterparts based on direct experiences, and these opinions can then be aggregated to build trust among unknown users. The mechanism to build new trust relationships based on existing ones is referred to as {\em trust inference}. State-of-the-art trust inference approaches employ the {\em transitivity} property of trust by propagating trust along connected users.
In this paper, we propose a novel trust inference model ({\em MaTrust}) by exploring an equally important property of trust, i.e., the {\em multi-aspect} property. MaTrust directly characterizes multiple {\em latent} factors for each trustor and trustee from the locally-generated trust relationships. Furthermore, it can naturally incorporate prior knowledge as {\em specified} factors. These factors in turn serve as the basis to infer the unseen trustworthiness scores. Experimental evaluations on real data sets show that the proposed MaTrust {\em significantly} outperforms several benchmark trust inference models in both effectiveness and efficiency.

\end{abstract}




\section{Introduction}
Trust is essential to reduce uncertainty and boost collaborations in many real-world applications including e-commerce~\cite{josang2002beta}, peer-to-peer networks~\cite{kamvar2003eigentrust}, semantic Web~\cite{richardson2003trust}, etc. In these applications, trust inference is widely used as the mechanism to build trust among unknown users. Typically, trust inference takes as input the existing trust ratings locally generated through direct interactions, and outputs an estimated trustworthiness score from a trustor to an unknown trustee. This trustworthiness score indicates to what degree the trustor could expect the trustee to perform a given action.

The basic assumption behind most of the existing trust inference methods is the {\em transitivity} property of trust~\cite{liu2011trust}, which basically means that if Alice trusts Bob and Bob trusts Carol, Alice might also trust Carol to some extent. These methods (see Section~\ref{sec:relatedwork} for a review), referred to as \emph{trust propagation} models as a whole, have been widely studied and successfully applied in many real-world settings~\cite{guha2004propagation,ziegler2005propagation,liu2011trust,kuter2007sunny,hang2009operators,massa2005controversial}.

In addition to transitivity, a few trust inference models explore another equally important property, the {\em multi-aspect} of trust~\cite{gefen2002reflections,sirdeshmukh2002consumer}. The basic assumption behind the multi-aspect methods is that trust is the composition of multiple factors, and different users may have different preferences for these factors. For example, in e-commerce, some users might care more about the factor of delivering time, whereas others give more weight to the factor of product price. However, the existing multi-aspect trust inference methods~\cite{sabater2002reputation,xiong2004peertrust,wang2011multi,tang2012mtrust} require as input more information (e.g., the delivering time as well as user's preference for it) in addition to locally-generated trust ratings, and therefore become infeasible in many trust networks where such information is not available.

Another limitation in existing trust inference models is that they tend to ignore some important prior knowledge during the inference procedure. In social science community, it is commonly known that {\em trust bias} is an integrated part in the final trust decision~\cite{tversky1974judgment}. Therefore, it would be helpful if we can incorporate such prior knowledge into the trust inference model. In computer science community, researchers begin to realize the importance of trust bias, and a recent work~\cite{mishra2011finding} models trustor bias as the propensity of a trustor to trust others.

In this paper, we focus on improving the trust inference accuracy by integrating the multi-aspect property and trust bias together, and the result is the proposed trust inference model \emph{MaTrust}. Different from the existing multi-aspect trust inference methods, MaTrust {\em directly} characterizes multiple latent factors for each trustor and trustee from existing trust ratings. In addition, MaTrust can naturally incorporate the priori knowledge (e.g., trust bias) as several specified factors. In particular, we consider three types of trust bias, i.e., {\em global bias}, {\em trustor bias}, and {\em trustee bias}. We will refer to the characterized latent and specified factors as {\em stereotypes} in the following. Finally, the characterized stereotypes are in turn used to estimate the trust ratings between unknown users. Compared with the existing multi-aspect methods, the proposed method is more general since it does not require any information as input other than the locally-generated trust ratings. Compared with
the trust propagation methods, our experimental evaluations on real data sets indicate that the proposed MaTrust is {\em significantly} better in both effectiveness and efficiency.


\hide{
To summarize, our main contributions are as follows:
\begin{itemize}
  \item A novel trust inference model MaTrust which captures the multi-aspect property of trust as well as the trust bias. Compared with the existing multi-aspect methods, the proposed method is more general since it does not require any information as input other than the locally-generated trust relationships.
  \item Compared with the several trust propagation methods, our experimental evaluations on real data sets indicate that the proposed MaTrust is {\em significantly} better in both effectiveness and efficiency.
\end{itemize}
}

The rest of the paper is organized as follows. Section~2 reviews related work. Section~3 presents the definition of the trust inference problem. Section~4 describes our optimization formulation for the problem defined in the previous section and shows how to incorporate the priori knowledge. Section~5 presents the inference algorithm to solve the formulation. Section~6 provides experimental results. Section~7 concludes the paper.

\section{Related Work}\label{sec:relatedwork}
In this section, we introduce related trust inference models including trust propagation models, multi-aspect trust inference models, and other related methods.

\subsection{Trust Propagation Models}

To date, a large body of trust inference models are based on trust propagation where trust is propagated along connected users in the trust network, i.e., the web of locally-generated trust ratings. Based on the interpretation of trust propagation, we further categorize these models into two classes: \emph{path interpretation} and \emph{component interpretation}.

In the first category of path interpretation, trust is propagated along a path from the trustor to the trustee, and the propagated trust from multiple paths can be combined to form a final trustworthiness score. For example, Wang and Singh~\cite{wang2006trust,wang2007formal} as well as Hang et al.~\cite{hang2009operators} propose operators to concatenate trust along a path and aggregate trust from multiple paths. Liu et al.~\cite{liu2010optimal} argue that not only trust values but social relationships and recommendation role are important for trust inference. In contrast, there is no explicit concept of paths in the second category of component interpretation. Instead, trust is treated as random walks on a graph or on a Markov chain~\cite{richardson2003trust}. Examples of this category include~\cite{guha2004propagation,massa2005controversial,ziegler2005propagation,kuter2007sunny,nordheimer2010trustworthiness}.

Different from these existing trust propagation models, the proposed MaTrust focuses on the multi-aspect of trust and directly characterizes several factors/aspects from the existing trust ratings. Compared with trust propagation models, our MaTrust has several unique advantages, including (1) multi-aspect property of trust can be captured; and (2) various types of prior knowledge can be naturally incorporated. In addition, one known problem about these propagation models is the slow on-line response speed~\cite{yao2012subgraph}, while MaTrust enjoys the {\em constant} on-line response time and the {\em linear} scalability for pre-computation.


\subsection{Multi-Aspect Trust Inference Models}

Researchers in social science have explored the multi-aspect property of trust for several years~\cite{sirdeshmukh2002consumer}. In computer science, there also exist a few trust inference models that {\em explicitly} explore the multi-aspect property of trust. For example, Xiong and Liu~\cite{xiong2004peertrust} model the value of the transaction in trust inference; Wang and Wu~\cite{wang2011multi} take competence and honesty into consideration; Tang et al.~\cite{tang2012mtrust} model aspect as a set of products that are similar to each other under product review sites; Sabater and Sierra~\cite{sabater2002reputation} divide trust in e-commerce environment into three aspects: price, delivering time, and quality.

However, all these existing multi-aspect trust inference methods require more information (e.g., value of transaction as well as user's preference for it, product and its category, etc.) and therefore become infeasible when such information is not available. In contrast, MaTrust does not require any information other than the locally-generated trust ratings, and could therefore be used in more general scenarios.

In terms of trust bias, Mishra et al.~\cite{mishra2011finding} propose an iterative algorithm to compute trustor bias. In contrast, our focus is to incorporate various types of trust bias as specified factors/aspects to increase the accuracy of trust inference.

\subsection{Other Related Methods}

Recently, researchers begin to apply machine learning models for trust inference. Nguyen et al.~\cite{nguyen2009trust} learn the importance of several trust-related features derived from a social trust framework. Our method takes a further step here by simultaneously learning the latent factors and the importance of bias. Seemingly similar concept of stereotype for trust inference is also used by Liu et al.~\cite{liu2009stereotrust} and Burnett et al.~\cite{burnett2010bootstrapping}. These methods learn the stereotypes from the user profiles of the trustees that the trustor has interacted with, and then use these stereotypes to reflect the trustor's first impression about unknown trustees. In contrast, MaTrust builds its stereotypes based on the existing trust ratings to capture multiple aspects for trust inference. There are also some recent work on using link prediction approaches to predict the {\em binary} trust/distrust relationship~\cite{leskovec2010predicting,chiang2011exploiting,hsieh2012low}. In this paper, we focus on the more general case where we want to infer a {\em continuous} trustworthiness score from the trustor to the trustee.

Finally, multi-aspect methods have been extensively studied in recommender systems~\cite{bell2007modeling,koren2009matrix,ma2009learning}. In terms of methodology, the closest related work is the collaborative filtering algorithm in~\cite{koren2009matrix}, which can be viewed as a special case of the proposed MaTrust. As mentioned before, our MaTrust is more general by learning the optimal weights for the prior knowledge and it leads to further performance improvement. On the application side, the goal of recommender systems is to predict users' flavors of items. It is interesting to point out that (1) on one hand, trust between users could help to predict the flavors as we may give more weight to the recommendations provided by trusted users; (2) on the other hand, trust itself might be affected by the similarity of flavors since users usually trust others with a similar taste~\cite{golbeck2009trust}. Although out of the scope of this paper, using recommendation to further improve trust inference accuracy might be an interesting topic for future work.

\section{Problem Definition}
In this section, we formally define our multi-aspect trust inference problem. Table~\ref{T:symbols} lists the main symbols we use throughout the paper.

\begin{table}[!h]
\caption{Symbols.}
\label{T:symbols}\centering
\begin{tabular}{|l|l|}
  \hline
  Symbol & Definition and Description \\ \hline \hline
  $\textbf{T}$ & the partially observed trust matrix \\
  $\textbf{F}$ & the characterized trustor matrix \\
  $\textbf{G}$ & the characterized trustee matrix \\
  $\textbf{F}_0$ & the sub-matrix of $\textbf{F}$ for latent factors \\
  $\textbf{G}_0$ & the sub-matrix of $\textbf{G}$ for latent factors \\
  $\textbf{T}'$ & the transpose of matrix $\textbf{T}$ \\
  $\textbf{T}(i,j)$ & the element at the $i^{th}$ row and $j^{th}$ column\\
   & of matrix $\textbf{T}$ \\
  $\textbf{T}(i,:)$ & the $i^{th}$ row of matrix $\textbf{T}$ \\
  $\textbf{T}(i,:)'$ & the transpose of vector $\textbf{T}(i,:)$ \\ \hline
  $\mathcal{K}$ & the set of observed trustor-trustee pairs in $\textbf{T}$ \\
  $\mu$ & the global bias \\
  $\textbf{x}$ & the vector of trustor bias \\
  $\textbf{y}$ & the vector of trustee bias \\
  $\textbf{x}(i)$ & the $i^{th}$ element of vector $\textbf{x}$ \\
  $n$ & the number of users \\
  $c$ & the number of specified factors \\
  $r$ & the number of latent factors \\
  $s$ & the number of all factors, $s=c+r$ \\
  $\alpha_1, \alpha_2, \alpha_3$ & the coefficients for specified factors \\
  $u$ & the trustor \\
  $v$ & the trustee \\
  $m_1, m_2$ & the maximum iteration number \\
  $\xi_1, \xi_2$ & the threshold to stop the iteration \\ \hline
\end{tabular}
\end{table}

Following conventions, we use bold capital letters for matrices, and bold lower case letters for vectors. For example, we use a partially observed matrix $\textbf{T}$ to model the locally-generated trust relationships, where the existing/observed trust relationships are represented as non-zero trust ratings and non-existing/unobserved relationships are represented as `?'. As for the observed trust rating, we represent it as a real number between 0 and 1 (a higher rating means more trustworthiness). We use calligraphic font $\mathcal{K}$ to denote the set of observed trustor-trustee indices in $\textbf{T}$. Similar to Matlab, we also denote the $i^{th}$ row of matrix $\textbf{T}$ as $\textbf{T}(i,:)$, and the transpose of a matrix with a prime. In addition, we denote the number of users as $n$ and the number of characterized factors as $s$.
Without loss of generality, we assume that the goal of our trust model is to infer the unseen trust relationship from the user $u$ to another user $v$, where $u$ is the trustor and $v$ is the unknown trustee to $u$.

Based on these notations, we first define the basic trust inference problem as follows:
\begin{problem}{The Basic Trust Inference Problem}\label{P:basic}
\begin{description}
\item[Given:] an $n \times n$ partially observed trust matrix $\textbf{T}$, a trustor $u$, and a trustee $v$, where $1 \leqslant u, v \leqslant n$ ($u\neq v$) and $\textbf{T}(u,v)$ = `?';
    \item [Find:] the estimated trustworthiness score $\mat{\hat T}(u,v)$.
    \end{description}
\end{problem}

In the above problem definition, given a trustor-trustee pair, the only information we need as input is the locally-generated trust ratings (i.e., the partially observed matrix $\textbf{T}$).
The goal of trust inference is to infer the new trust ratings (i.e., unseen/unobserved trustworthiness scores in the partially observed matrix $\textbf{T}$) by collecting the knowledge from existing trust relationships. In this paper, we assume that we can access such existing trust relationships. For instance, these relationships could be collected by central servers in a centralized environment like eBay, or by individuals in a distributed environment like EigenTrust~\cite{kamvar2003eigentrust}. How to collect these trust relationships is out of the scope of this work.

In this paper, we propose a multi-aspect model for such trust inference in Problem~\ref{P:basic}. That is, we want to infer an $n \times s$ {\em trustor matrix} $\mat F$ whose element indicates to what extent the corresponding person trusts others wrt a specific aspect/factor. Similarly, we want to infer another $n \times s$ {\em trustee matrix} $\mat G$ whose element indicates to what extent the corresponding person is trusted by others wrt a specific aspect/factor. Such trustor and trustee matrices are in turn used to infer the unseen trustworthiness scores.  Based on the basic trust inference problem, we define the multi-aspect trust inference problem under MaTrust as follows:
\begin{problem}{The MaTrust Trust Inference Problem}\label{P:matrust}
\begin{description}
\item[Given:] an $n \times n$ partially observed trust matrix $\textbf{T}$, the number of factors $s$, a trustor $u$, and a trustee $v$, where $1 \leqslant u, v \leqslant n$ ($u\neq v$) and $\textbf{T}(u,v)$ = `?';
    \item [Find:] (1) an $n \times s$ trustor matrix $\mat F$ and an $n \times s$ trustee matrix $\mat G$; (2) the estimated trustworthiness score $\mat{\hat T}(u,v)$.
    \end{description}
\end{problem}

\subsection{An Illustrative Example}

\begin{figure*}[!t]
\centering
  \subfigure[The observed locally-generated pair-wise trust relationships]{
    \label{F:example:events}\centering
    \includegraphics[width=1.5in]{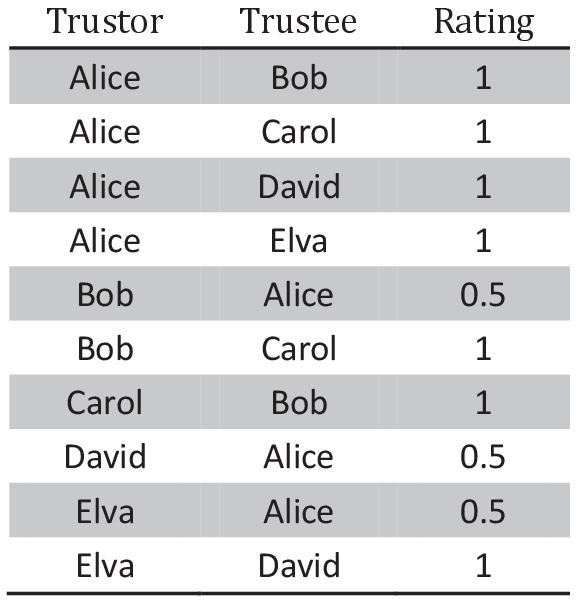}}
  \hspace{0.1in}
  \subfigure[The partially observed trust matrix \textbf{T}]{
    \label{F:example:t}\centering
    \includegraphics[width=2.0in]{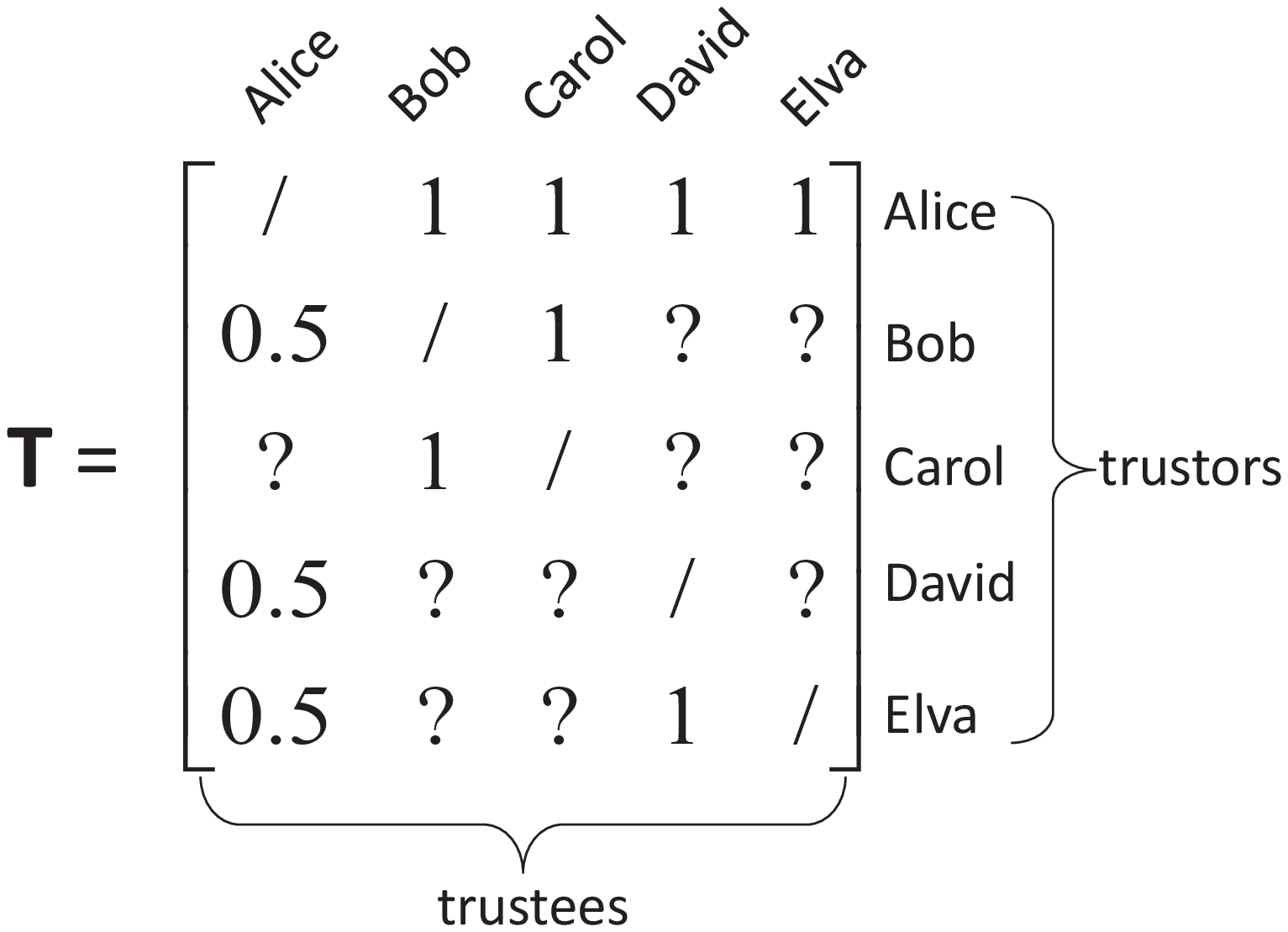}}
    \hspace{0.1in}
  \subfigure[The inferred trustor matrix \textbf{F} and trustee matrix \textbf{G}]{
    \label{F:example:fg}\centering
    \includegraphics[width=3.0in]{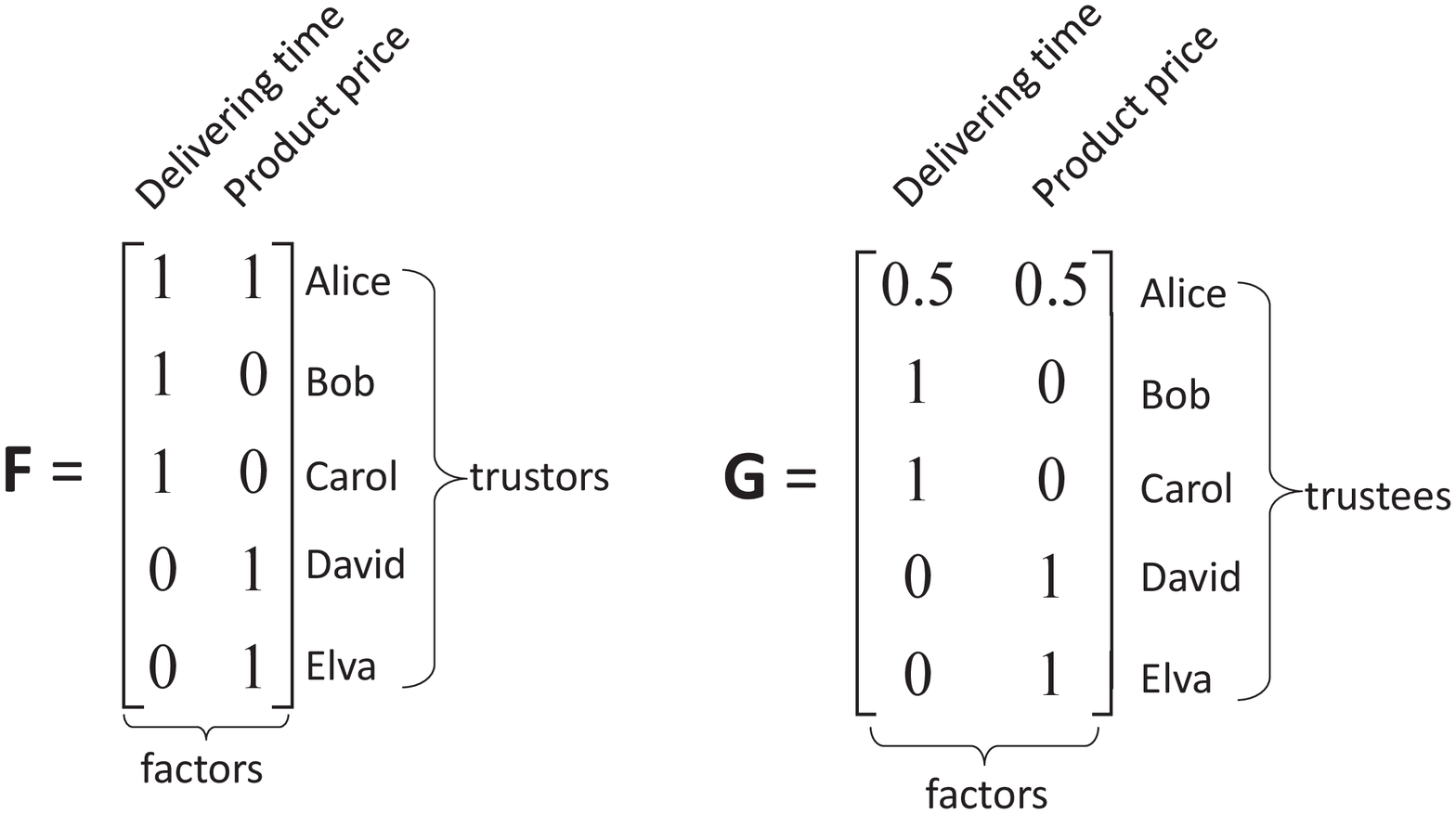}}
\caption{An illustrative example for MaTrust.}
\label{F:example}\centering
\end{figure*}

To further illustrate our MaTrust trust inference problem (Problem~\ref{P:matrust}), we give an intuitive example as shown in Fig.~\ref{F:example}.

\hide{
\begin{eqnarray}
{\mat{T}=\begin{pmatrix}
            / & 1 & 1 & 1 & 1\\
            0.5 & / & 1 & ? & ?\\
            ? & 1 & / & ? & ?\\
            0.5 & ? & ? & / & ?\\
            0.5 & ? & ? & 1 & /\\
                        \end{pmatrix} }
                      \nonumber  \\
{\mat{F}=\begin{pmatrix}
            1 & 1 \\
            1 & 0 \\
            1 & 0 \\
            0 & 1 \\
            0 & 1 \\
  \end{pmatrix} } ~~~
{\mat{G}=\begin{pmatrix}
            0.5 & 0.5 \\
            1 & 0 \\
            1 & 0 \\
            0 & 1 \\
            0 & 1 \\
                        \end{pmatrix} }                      \nonumber
\end{eqnarray}
}

In this example, we observe several locally-generated pair-wise trust relationships between five users (e.g., `{\em Alice}', `{\em Bob}', `{\em Carol}', `{\em David}', and `{\em Elva}') as shown in Fig.~\ref{F:example:events}. Each observation contains a trustor, a trustee, and a numerical trust rating from the trustor to the trustee. We then model these observations as a $5 \times 5$ partially observed matrix $\textbf{T}$ (see Fig.~\ref{F:example:t}) where $\textbf{T}(i,j)$ is the trust rating from the $i^{th}$ user to the $j^{th}$ user if the rating is observed and $\textbf{T}(i,j)$ = `?' otherwise. Notice that we do not consider self-ratings and thus represent the diagonal elements of $\textbf{T}$ as `$/$'. By setting the number of factors $s = 2$, our goal is to infer two $5 \times 2$ matrices $\textbf{F}$ and $\textbf{G}$ (see Fig.~\ref{F:example:fg}) from the input matrix $\textbf{T}$. Each row of the two matrices is the stereotype for the corresponding user, and each column of the matrices represents a certain aspect/factor in trust inference (e.g., `delivering time', `product price', etc).
For example, we can see that {\em Alice} trusts others strongly wrt both `delivering time' and `product price' (based on matrix $\mat F$), and she is in turn moderately trusted by others wrt these two factors (based on matrix $\mat G$). On the other hand, both {\em Bob} and {\em Carol} put more emphasis on the delivering time, while {\em David} and {\em Elva} care more about the product price.

Once $\textbf{F}$ and $\textbf{G}$ are inferred, we can use these two matrices to estimate the unseen trustworthiness scores (i.e., the `?' elements in $\textbf{T}$). For instance, the trustworthiness from  {\em Carol} to {\em Alice} can be estimated as $\mat{\hat T}(3,1) = \textbf{F}(3,:) \textbf{G}(1,:)' = 0.5$. This estimation is reasonable because Carol has the same stereotype as Bob and the trustworthiness score from {\em Bob} to {\em Alice} is also $0.5$. As another example, {\em David} and {\em Elva} have similar preferences (i.e., the same stereotypes), and thus we conjecture that they would trust each other (i.e., $\mat{\hat T}(4,5)$ = $\textbf{F}(4,:) \textbf{G}(5,:)' = 1$). In the rest of the paper, we will mainly focus on how to  characterize $\textbf{F}$ and $\textbf{G}$ from the partially observed input matrix $\mat T$.

\section{The Proposed Optimization Formulation}
In this section, we present our optimization formulation for the problem defined in the previous section. We start with the basic form, and then show how to incorporate the trust bias as specified factors followed by its equivalent formulation. Finally, we discuss some generalizations of our formulation.


\subsection{The Basic Formulation}

Formally, Problem~\ref{P:matrust} can be formulated as the following optimization problem: \begin{eqnarray}\label{E:optbasic}
\min_{\textbf{F},\textbf{G}} \sum_{(i,j)\in \mathcal{K}} (\textbf{T}(i,j) - \textbf{F}(i,:)\textbf{G}(j,:)')^2 + \lambda ||\textbf{F}||_{fro}^2 + \lambda ||\textbf{G}||_{fro}^2
\end{eqnarray}
\noindent where $\lambda$ is a regularization parameter; $||\textbf{F}||_{fro}$ and $||\textbf{G}||_{fro}$ are the Frobenious norm of the trustor and trustee matrices, respectively.

By this formulation, MaTrust aims to minimize the squared error on the set of observed trust ratings. Notice that in Eq.~\eqref{E:optbasic}, we have two additional regularization terms ($||\textbf{F}||_{fro}^2$ and $||\textbf{G}||_{fro}^2$) to improve the solution stability. The parameter $\lambda \ge 0$ controls the amount of such regularizations. Based on the resulting $\textbf{F}$ and $\textbf{G}$ of the above equation, the unseen trustworthiness score $\mat{\hat T}(u,v)$ can then be estimated by the corresponding stereotypes $\textbf{F}(u,:)$ and $\textbf{G}(v,:)$ as: \begin{equation}
\mat{\hat T}(u,v) = \textbf{F}(u,:)\textbf{G}(v,:)'
\end{equation}

\subsection{Incorporating Bias}

The above formulation can naturally incorporate some prior knowledge such as trust bias into the inference procedure. In this paper, we explicitly consider the following three types of trust bias: \emph{global bias}, \emph{trustor bias}, and \emph{trustee bias}, although other types of bias can be incorporated in a similar way.
\begin{itemize}
  \item The {\em global bias} represents the average level of trust in the community. The intuition behind this is that users tend to rate optimistically in some reciprocal environments (e.g., e-Commerce~\cite{resnick2002trust}) while they are more conservative in others (e.g., security-related applications). As a result, it might be useful to take such global bias into account and we model it as a scalar $\mu$.
  \item The {\em trustor bias} is based on the observation that some trustors tend to generously give higher trust ratings than others. This bias reflects the propensity of a given trustor to trust others, and it may vary a lot among different trustors. Accordingly, we can model the trustor bias as vector \textbf{x} with $\textbf{x}(i)$ indicating the trust propensity of the $i^{th}$ trustor.
  \item The third type of bias ({\em trustee bias}) aims to characterize the fact that some trustees might have relatively higher capability in terms of being trusted than others. Similar to the second type of bias, we model this type of bias as vector \textbf{y}, where $\textbf{y}(j)$ indicates the overall capability of the $j^{th}$ trustee compared to the average.
\end{itemize}

Each of these three types of bias can be represented as a {\em specified} factor for our MaTrust model, respectively. By incorporating such bias into Eq.~\eqref{E:optbasic}, we have the following formulation: \begin{eqnarray}\label{E:optbias}
\min_{\textbf{F},\textbf{G}} \sum_{(i,j)\in \mathcal{K}} &~& (\textbf{T}(i,j) - \textbf{F}(i,:)\textbf{G}(j,:)')^2 + \lambda ||\textbf{F}||_{fro}^2 + \lambda ||\textbf{G}||_{fro}^2\nonumber\\
\textrm{Subject~to:~}&~& \mat{F}(:,1) = \mu\mat{1},~\mat{G}(:,1)=\alpha_1\mat{1}~~~\textrm{(global~bias)}\nonumber\\
&~& \mat{F}(:,2) = \mat{x},~\mat{G}(:,2)=\alpha_2\mat{1}~~~\textrm{(trustor~bias)}\nonumber\\
&~& \mat{F}(:,3) = \alpha_3\mat{1},~\mat{G}(:,3)=\mat{y}~~~\textrm{(trustee~bias)}\end{eqnarray}
\noindent where $\alpha_1,\alpha_2,$ and $\alpha_3$ are the weights of bias that we need to estimate based on the existing trust ratings.

In addition to these three {\em specified} factors, we refer to the remaining factors in the trustor and trustee matrices as {\em latent} factors. To this end, we define two $n \times r$ sub-matrices of $\mat{F}$ and $\mat{G}$ for the latent factors. That is, we define $\mat{F}_0 = \mat{F}(: , 4:s)$ and $\mat{G}_0 = \mat{G}(: , 4:s)$, where each column of $\mat{F}_0$ and $\mat{G}_0$ corresponds to one latent factor and $r$ is the number of latent factors. With this notation, we have the following equivalent form of Eq.~\eqref{E:optbias}: \begin{eqnarray}\label{E:optfinal}
\min_{\textbf{F}_0,\textbf{G}_0,\alpha_1,\alpha_2,\alpha_3} \sum_{(i,j)\in \mathcal{K}}&~& (\textbf{T}(i,j) - (\alpha_1 \mu + \alpha_2 \textbf{x}(i) + \alpha_3 \textbf{y}(j) + \nonumber\\ &~& \textbf{F}_0(i,:)\textbf{G}_0(j,:)'))^2 + \lambda\|\mat{F}\|_{fro}^2  + \lambda\|\mat{G}\|_{fro}^2
\end{eqnarray}

Notice that there is no coefficient before $\textbf{F}_0(i,:)\textbf{G}_0(j,:)'$ as it will be automatically absorbed into $\textbf{F}_0$ and $\textbf{G}_0$. Once we have inferred all the parameters (i.e., $\textbf{F}_0$, $\textbf{G}_0$, $\alpha_1$, $\alpha_2$, and $\alpha_3$) of Eq.~\eqref{E:optfinal}, the unseen trustworthiness score $\mat{\hat T}(u,v)$ can be immediately estimated as: \begin{equation}\label{E:onlineinfer}
\mat{\hat T}(u,v) = \textbf{F}_0(u,:)\textbf{G}_0(v,:)' + \alpha_1 \mu + \alpha_2 \textbf{x}(u) + \alpha_3 \textbf{y}(v)
\end{equation}

In the above formulations, we need to specify the three types of bias, i.e., to compute $\mu$, $\textbf{x}$, and $\textbf{y}$. Remember that the only information we need for MaTrust is the existing trust ratings. Therefore, we simply estimate the bias information from $\textbf{T}$ as follows: \begin{equation}\label{E:bias}
\left\{ \begin{aligned}
         \mu = \sum_{(i,j)\in \mathcal{K}} \textbf{T}(i,j) / |\mathcal{K}|~~~~~~~~~~~~~~~\\
         \textbf{x}(i) = \sum_{j,(i,j)\in \mathcal{K}} \textbf{T}(i,j)/|row_i| - \mu \\
         \textbf{y}(j) = \sum_{i,(i,j)\in \mathcal{K}} \textbf{T}(i,j)/|col_j| - \mu \\
          \end{aligned} \right.
\end{equation}
where $|row_i|$ is the number of the observed elements in the $i^{th}$ row of $\textbf{T}$, and $|col_j|$ is the number of the observed elements in the $j^{th}$ column of $\textbf{T}$.

\subsection{Discussions and Generalizations}

We further present some discussions and generalizations of our optimization formulation.

First, it is worth pointing out that our formulation in Eq.~\eqref{E:optbasic} differs from the standard matrix factorization (e.g., SVD) as in the objective function, we try to minimize the square loss {\em only} on those observed trust pairs. This is because the majority of trust pairs are missing from the input trust matrix $\mat T$. In this sense, our problem setting is conceptually similar to the standard collaborative filtering, as in both cases, we aim to fill in missing values in a partially observed matrix (trustor-trustee matrix vs. user-item matrix). Indeed, if we fix the coefficients $\alpha_1=\alpha_2=\alpha_3=1$ in Eq.~\eqref{E:optbias}, it is reduced to the collaborative filtering algorithm in~\cite{koren2009matrix}. Our formulation in Eq.~\eqref{E:optbias} is more general as it also allows to learn the optimal coefficients from the input trust matrix $\mat T$. Our experimental evaluations show that such subtle treatment is crucial and it leads to further performance improvement over these existing techniques.

Second, although our MaTrust is a subjective trust inference metric where different trustors may form different opinions on the same trustee~\cite{massa2005controversial}, as a side product, the proposed MaTrust can also be used to infer an objective, unique trustworthiness score for each trustee. For example, this objective trustworthiness score can be computed based on the trustee matrix $\textbf{G}$. We will compare this feature of MaTrust with a well studied objective trust inference metric EigenTrust~\cite{kamvar2003eigentrust} in the experimental evaluation section.

Finally, we would like to point out that our formulation is flexible and can be generalized to other settings. For instance, our current formulation adopts the square loss function in the objective function. In other words, we implicitly assume that the residuals of the pair-wise trustworthiness scores follow a Gaussian distribution, and in our experimental evaluations, we found it works well. Nonetheless, our upcoming proposed MaTrust algorithm can be generalized to {\em any} Bregman divergence in the objective function. Also, we can naturally incorporate some additional constraints (e.g., non-negativity, sparseness, etc) in the trustor and trustee matrices. After we infer all the parameters (e.g., the coefficients for the bias, and the trustor and trustee matrices, etc), we use a linear combination (i.e., inner product) of the trustor stereotype (i.e., $\textbf{F}(u,:)$) and trustee stereotype (i.e., $\textbf{G}(v,:)$) to compute the trustworthiness score $\mat{\hat T}(u,v)$. We can also generalize this linear form to other non-linear combinations, such as the logistic function. For the sake of clarity, we skip the details of such generalizations in the paper.


\hide{
We now discuss some generations to our basic formulation.

First, in our basic formulation, we use linear combination (i.e., inner product) of the trustor stereotype (i.e., $\textbf{F}(u,:)$) and trustee stereotype (i.e., $\textbf{G}(v,:)$) to compute the trustworthiness score $\mat{\hat T}(u,v)$. Other non-linear combinations, such as logistic function, can also be used to combine these two vectors.

Second, the existing trust relationships are usually sparse in many real-world settings~\cite{guha2004propagation,golbeck2006inferring}. Therefore, we can store them in the sparse format which could save a lot of space for MaTrust.

Finally, in addition to bias, our formulation allows the incorporation of other additional knowledge such as temporal dynamics and confidence levels. We leave it as future work.
}

\section{The Proposed MaTrust Algorithm}
In this section, we present the proposed algorithm to solve the MaTrust trust inference problem (i.e., Eq.~\eqref{E:optfinal}), followed by some effectiveness and efficiency analysis.

\subsection{The MaTrust Algorithm}

Unfortunately, the optimization problem in Eq.~\eqref{E:optfinal} is not jointly convex wrt the coefficients ($\alpha_1$, $\alpha_2$, and $\alpha_3$) and the trustor/trustee matrices ($\textbf{F}_0$ and $\textbf{G}_0$) due to the coupling between them. Therefore, instead of seeking for a global optimal solution, we try to find a local minima by alternatively updating the coefficients and the trustor/trustee matrices while fixing the other.
The alternating procedure will lead to a local optima when the convergence criteria are met, i.e., either the $L_2$ norm between successive estimates of both $\textbf{F}$ and $\textbf{G}$ (which are equivalent to $\alpha_1$, $\alpha_2$, $\alpha_3$, $\textbf{F}_0$, and $\textbf{G}_0$) is below our threshold $\xi_1$ or the maximum iteration step $m_1$ is reached.


\subsubsection{Sub-routine 1: updating the trustor/trustee matrices}

First, let us consider how to update the trustor/trustee matrices ($\textbf{F}_0$ and $\textbf{G}_0$) when we fix the coefficients ($\alpha_1$, $\alpha_2$, and $\alpha_3$). For clarity, we define an $n \times n$ matrix $\textbf{P}$ as follows: \begin{equation}\label{E:t2p}
\textbf{P}(i,j) = \left\{ \begin{array}{ll}
\textbf{T}(i,j) - (\alpha_1 \mu + \alpha_2 \textbf{x}(i) + \alpha_3 \textbf{y}(j)) & \textrm{if $(i,j)\in {\cal {K}}$}\\
\textrm{`?'} & \textrm{otherwise}
\end{array} \right.
\end{equation}
where $\alpha_1$, $\alpha_2$, and $\alpha_3$ are some fixed constants.

\hide{
\begin{eqnarray}\label{E:t2p}
\textbf{P}(i,j) &=& {\textbf{T}(i,j) - (\alpha_1 \mu + \alpha_2 \textbf{x}(i) + \alpha_3 \textbf{y}(j))~\textrm{if~} (i,j)\in {\cal {K}}}
\nonumber\\
~&=& \textrm{`?'}~\textrm{otherwise}
\end{eqnarray}
}

Based on the above definition, Eq.~\eqref{E:optfinal} can be simplified (by ignoring some constant terms) as: \begin{equation}\label{E:optsimple}
\min_{\textbf{F}_0,\textbf{G}_0} \sum_{(i,j)\in \mathcal{K}} (\textbf{P}(i,j) - \textbf{F}_0(i,:)\textbf{G}_0(j,:)')^2 + \lambda ||\textbf{F}_0||_{fro}^2 + \lambda ||\textbf{G}_0||_{fro}^2
\end{equation}

Therefore, updating the trustor/trustee matrices when we fix the coefficients unchanged becomes a standard matrix factorization problem for missing values. Many existing algorithms (e.g.,~\cite{koren2009matrix,ma2009learning,buchanan2005damped}) can be plugged in to solve Eq.\eqref{E:optsimple}. In our experiment, we found the so-called alternating strategy, where we recursively update one of the two trustee/trustor matrices while keeping the other matrix fixed, works best and thus recommend it in practice. A brief skeleton of the algorithm is shown in Algorithm~\ref{A:skeleton}, and the detailed algorithms are presented in the appendix for completeness.


\begin{algorithm}[t]
\caption{updateMatrix(\textbf{P}, $r$). (See the appendix for the details)}\label{A:skeleton}
\begin{algorithmic}[1]
  \REQUIRE {The $n \times n$ matrix $\textbf{P}$, and the latent factor size $r$}
  \ENSURE {The $n \times r$ trustor matrix $\textbf{F}_0$, and the $n \times r$ trustee matrix $\textbf{G}_0$}
  \STATE [$\textbf{F}_0$, $\textbf{G}_0$] = alternatingFactorization(\textbf{P}, $r$);
  \RETURN [$\textbf{F}_0$, $\textbf{G}_0$];
\end{algorithmic}
\end{algorithm}

\subsubsection{Sub-routine 2: updating the coefficients}

Here, we consider how to update the coefficients ($\alpha_1$, $\alpha_2$, and $\alpha_3$) when we fix the trustor/trustee matrices.

If we fix the trustor and trustee matrices ($\mat{F}_0$ and $\mat{G}_0$) and let: \begin{equation}\label{E:t2p2}
\textbf{P}(i,j) = \left\{ \begin{array}{ll}
\textbf{T}(i,j) - \textbf{F}_0(i,:)\textbf{G}_0(j,:)' & \textrm{if $(i,j)\in {\cal {K}}$}\\
\textrm{`?'} & \textrm{otherwise}
\end{array} \right.
\end{equation}
Eq.~\eqref{E:optfinal} can then be simplified (by dropping constant terms) as: \begin{eqnarray}\label{E:optsimplecoeff}
\min_{\alpha_1, \alpha_2, \alpha_3} \sum_{(i,j)\in \mathcal{K}} (\textbf{P}(i,j) - (\alpha_1 \mu + \alpha_2 \mat{x}(i) + \alpha_3 \mat{y}(j)))^2 + n\lambda\sum_{i=1}^3 \alpha_i^2
\end{eqnarray}

\hide{
\begin{eqnarray}\label{E:t2p2}
\textbf{P}(i,j) &=& {\textbf{T}(i,j) - \textbf{F}_0(i,:)\textbf{G}_0(j,:)'~\textrm{if~} (i,j)\in {\cal {K}}}\nonumber\\
               ~ &=& \textrm{`?'}~\textrm{otherwise}\end{eqnarray}
}

To simplify the description, let us introduce another scalar $k$ to index each pair $(i,j)$ in the observed trustor-trustee pairs $\cal K$, that is, $(i,j)\in{\cal{K}}\rightarrow k=\{1,2,...,|{\cal{K}}|\}$. Let $\mat b$ denote a vector of length $|\mathcal{K}|$ with $\mat{b}(k)=\textbf{P}(i,j)$. We also define a $|\mathcal{K}|\times 3$ matrix $\mat A$ as: $\textbf{A}(k,1) = \mu$, $\textbf{A}(k,2) = \textbf{x}(i)$, $\textbf{A}(k,3) = \textbf{y}(j)$ $(k=1,2,...,|\mathcal{K}|)$; and a $3 \times 1$ vector $\mat{\alpha}=[\alpha_1, \alpha_2, \alpha_3]'$. Then, Eq.~\eqref{E:optsimplecoeff} can be formulated as the ridge regression problem wrt the vector $\mat{\alpha}$:\begin{eqnarray}
\min_{\mat{\alpha}} ||\mat{b} - \mat{A}\mat{\alpha}||_2^2 + n\lambda\|\mat{\alpha}\|^2
\end{eqnarray}
In practice, we shrink the regularization parameter in the above equation from $n\lambda$ to $\lambda$ to strengthen the importance of bias. Therefore, we can update the coefficients as: \begin{equation}\label{E:linearregression}
\mat{\alpha}=[\alpha_1, \alpha_2, \alpha_3]' = (\textbf{A}' \textbf{A} + \lambda\mat{I}_{3 \times 3})^{-1} \textbf{A}' \textbf{b}
\end{equation}

\subsubsection{Putting everything together: MaTrust}

\begin{algorithm}[t]
\caption{MaTrust($\textbf{T}$, $\mathcal{K}$, $r$, $u$, $v$).}\label{A:matrust}
\begin{algorithmic}[1]
  \REQUIRE {The $n \times n$ partially observed trust matrix $\textbf{T}$, the set of observed trustor-trustee pairs $\mathcal{K}$, the latent factor size $r$, trustor $u$, and trustee $v$}
  \ENSURE {The estimated trustworthiness score $\mat{\hat T}(u,v)$}
  \STATE [$\mu$, \textbf{x}, \textbf{y}] $\leftarrow$ computeBias($\textbf{T}$);
  \STATE initialize $\alpha_1 = \alpha_2 = \alpha_3 = 1$;
  \WHILE {not convergent}
    \FOR {each $(i,j) \in \mathcal{K}$}
      \STATE $\textbf{P}(i,j)$ $\leftarrow$ $\textbf{T}(i,j) - (\alpha_1 \mu + \alpha_2 \textbf{x}(i) + \alpha_3 \textbf{y}(j))$;
    \ENDFOR
    \STATE [$\textbf{F}_0$, $\textbf{G}_0$] = updateMatrix($\textbf{P}$, $r$);
    \FOR {each $(i,j) \in \mathcal{K}$}
      \STATE $\textbf{P}(i,j)$ $\leftarrow$ $\textbf{T}(i,j) - \textbf{F}_0(i,:) \textbf{G}_0(j:,)'$;
    \ENDFOR
    \STATE $[\alpha_1, \alpha_2, \alpha_3]'$ = updateCoefficient($\textbf{P}$, $\mu$, \textbf{x}, \textbf{y});
  \ENDWHILE
  \RETURN $\mat{\hat T}(u,v) \leftarrow \textbf{F}_0(u,:)\textbf{G}_0(v,:)' + \alpha_1 \mu + \alpha_2 \textbf{x}(u) + \alpha_3 \textbf{y}(v)$;
\end{algorithmic}
\end{algorithm}


Putting everything together, we propose Algorithm~\ref{A:matrust} for our MaTrust trust inference problem. The algorithm first uses Eq.~\eqref{E:bias} to compute the global bias, trustor bias, and trustee bias (Step~1), and initializes the coefficients (Step~2). Then the algorithm begins the alternating procedure (Step~3-12). First, it fixes $\alpha_1$, $\alpha_2$, and $\alpha_3$, and applies Eq.~\eqref{E:t2p} to incorporate bias. After that, the algorithm invokes Algorithm~\ref{A:skeleton} to update the trustor matrix $\textbf{F}_0$ and trustee matrix $\textbf{G}_0$. Next, the algorithm fixes $\textbf{F}_0$ and $\textbf{G}_0$, and uses ridge regression in Eq.~\eqref{E:linearregression} to update $\alpha_1$, $\alpha_2$, and $\alpha_3$. The alternating procedure ends when the stopping criteria of Eq.~\eqref{E:optfinal} are met. Finally, the algorithm outputs the estimated trustworthiness from the given trustor $u$ to the trustee $v$  using Eq.~\eqref{E:onlineinfer} (Step~13).

It is worth pointing out that Step 1-12 in the algorithm can be pre-computed and their results ($\textbf{F}_0$, $\textbf{G}_0$, $\alpha_1$, $\alpha_2$, and $\alpha_3$) can be stored in the off-line/pre-computational stage. When an on-line trust inference request arrives, MaTrust only needs to apply Step 13 to return the inference result, which only requires a constant time.

\subsection{Algorithm Analysis}\label{sec:algorithmanalysis}

Here, we briefly analyze the effectiveness and efficiency of our MaTrust algorithm and provide detailed proofs in the appendix.

The effectiveness of the proposed MaTrust algorithm can be summarized in Lemma~\ref{L:effectiveness}, which says that overall, it finds a local minima solution. Given that the original optimization problem in Eq.~\eqref{E:optfinal} is not jointly convex wrt the coefficients ($\alpha_1$, $\alpha_2$, and $\alpha_3$) and the trustor/trustee matrices ($\textbf{F}_0$ and $\textbf{G}_0$), such a local minima is acceptable in practice.

\begin{lemma}{\bf {Effectiveness of MaTrust}}.\label{L:effectiveness}
Algorithm~\ref{A:matrust} finds a local minima for the optimization problem in Eq.~\eqref{E:optfinal}.
\end{lemma}
\proof See the Appendix \QED

The time complexity of the proposed MaTrust is summarized in Lemma~\ref{L:efficiency}, which says that MaTrust scales {\em linearly} wrt the number of users and the number of the observed trustor-trustee pairs.

\begin{lemma}{\bf {Time Complexity of MaTrust}}.\label{L:efficiency}
Algorithm~\ref{A:matrust} requires $O(nr^3m_1m_2 + |\mathcal{K}|r^2m_1m_2)$ time, where $m_1$ and $m_2$ are the maximum iteration numbers in Algorithm~\ref{A:matrust} and Algorithm~\ref{A:skeleton}, respectively.
\end{lemma}
\proof See the Appendix \QED

The space complexity of MaTrust is summarized in Lemma~\ref{L:efficiency2}, which says that MaTrust requires {\em linear} space wrt the number of users and the number of the observed trustor-trustee pairs.

\begin{lemma}{\bf {Space Complexity of MaTrust}}.\label{L:efficiency2}
Algorithm~\ref{A:matrust} requires $O(|\mathcal{K}| + nr + r^2)$ space.
\end{lemma}
\proof See the Appendix \QED

Notice that for both time complexity and space complexity, we have a {\em polynomial} term wrt the number of the latent factors $r$. In practice, this parameter is small compared with the number of the users ($n$) or the number of the observed trustor-trustee pairs ($|\mathcal{K}|$). For example, in our experiments, we did not observe significant performance improvement when the number of latent factors is larger than $10$ (See the next section for the detailed evaluations).

\section{Experimental Evaluation}\label{sec:exp}
\begin{table*}[!t]
\caption{High level statistics of advogato and PGP data sets.}
\label{T:statistics}\centering
\begin{tabular}{c||c|c|c|c|c|c}
  \hline
  Data set & Nodes & Edges & Avg. degree & Avg. clustering~\cite{watts1998collective} & Avg. diameter~\cite{leskovec2005graphs} & Date \\ \hline \hline
  advogato-1 & 279 & 2,109 & 15.1 & 0.45 & 4.62 & 2000-02-05 \\
  advogato-2 & 1,261 & 12,176 & 19.3 & 0.36 & 4.71 & 2000-07-18 \\
  advogato-3 & 2,443 & 22,486 & 18.4 & 0.31 & 4.67 & 2001-03-06 \\
  advogato-4 & 3,279 & 32,743 & 20.0 & 0.33 & 4.74 & 2002-01-14 \\
  advogato-5 & 4,158 & 41,308 & 19.9 & 0.33 & 4.83 & 2003-03-04 \\
  advogato-6 & 5,428 & 51,493 & 19.0 & 0.31 & 4.82 & 2011-06-23 \\
  PGP & 38,546 & 317,979 & 16.5 & 0.45 & 7.70 & 2008-06-05 \\ \hline
\end{tabular}
\end{table*}

In this section, we present experimental evaluations, after we introduce the data sets. All the experiments are designed to answer the following questions: \begin{itemize}
\item {\em Effectiveness}: How accurate is the proposed MaTrust for trust inference? How robust is the inference result wrt the different parameters in MaTrust?
\item {\em Effciency}: How fast is the proposed MaTrust? How does it scale?
\end{itemize}

\subsection{Data Sets Description}

Many existing trust inference models design specific simulation studies to verify the underlying assumptions of the corresponding inference models. In contrast, we focus on two widely used real, benchmark data sets in order to compare the performance of different trust inference models.


The first data set is {\em advogato}\footnote{http://www.trustlet.org/wiki/Advogato\_dataset.}. It is a trust-based social network for open source developers. To allow users to certify each other, the network provides 4 levels of trust assertions, i.e., `{\em Observer}', `{\em Apprentice}', `{\em Journeyer}', and `{\em Master}'. These assertions can be mapped into real numbers which represent the degree of trust. To be specific, we map `{\em Observer}', `{\em Apprentice}', `{\em Journeyer}', and `{\em Master}' to 0.1, 0.4, 0.7, and 0.9, respectively (a higher value means more trustworthiness).


The second data set is {\em PGP} (short for Pretty Good Privacy)~\cite{hang2009operators}. PGP adopts the concept of `web of trust' to establish a decentralized model for data encryption and decryption. Similar to {\em advogato}, the web of trust in PGP data set contains 4 levels of trust as well. In our experiments, we also map them to 0.1, 0.4, 0.7, and 0.9, respectively.

Table~\ref{T:statistics} summarizes the basic statistics of the two resulting partially observed trust matrices $\mat T$. Notice that for the {\em advogato} data set, it contains six different snapshots, i.e., {\em advogato-1}, {\em advogato-2},..., {\em advogato-6}, etc. We use the largest snapshot (i.e., {\em advogato-6}) in the following unless otherwise specified.

\begin{figure}[!t]
  \centering
  \subfigure[Trustor bias distribution on advogato]{
  \label{F:trustorbias:advogato}\centering
    \includegraphics[width=1.5in]{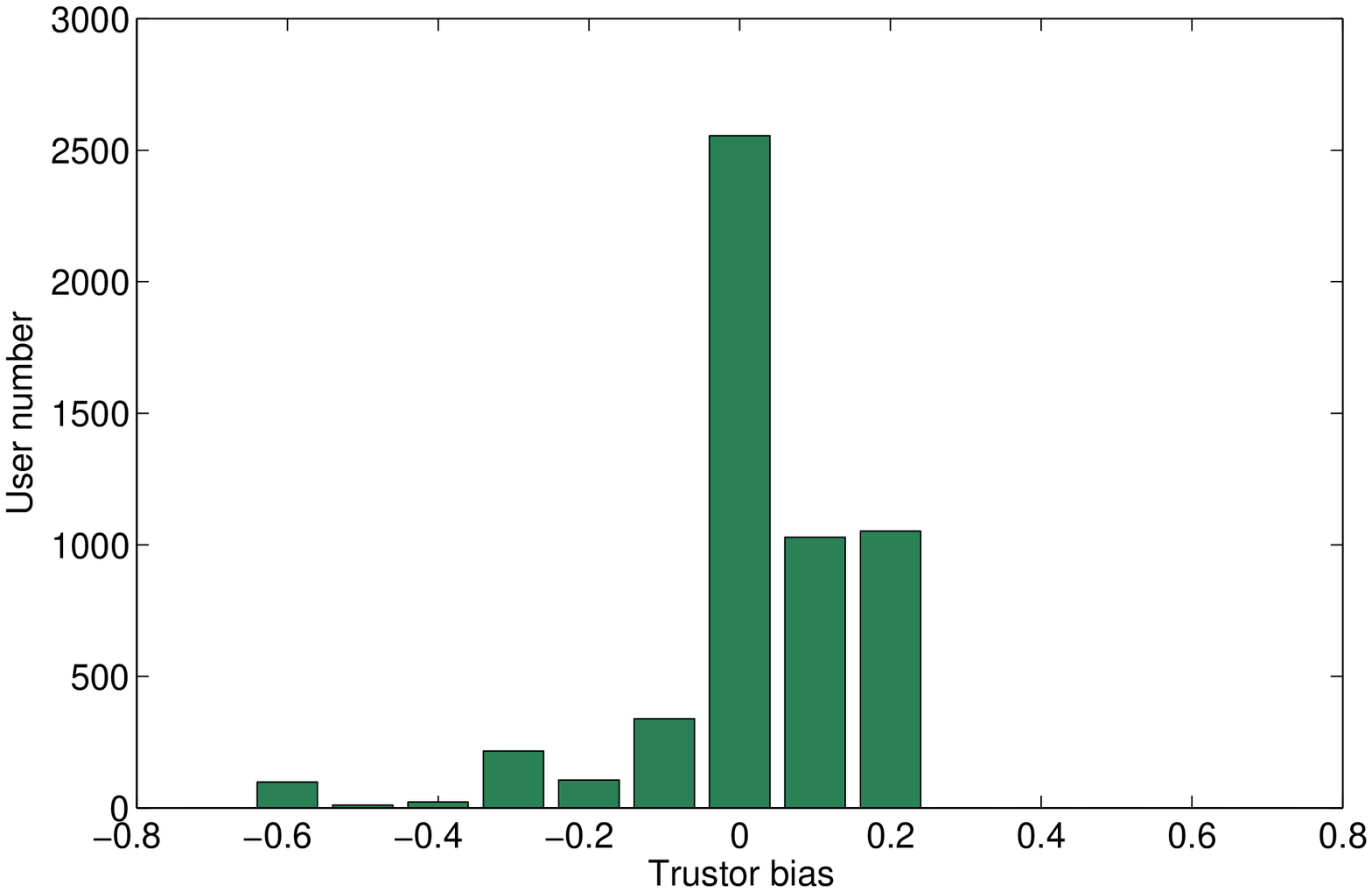}}
    \hspace{0.1in}
  \subfigure[Trustee bias distribution on advogato]{
  \label{F:trusteebias:advogato}\centering
    \includegraphics[width=1.5in]{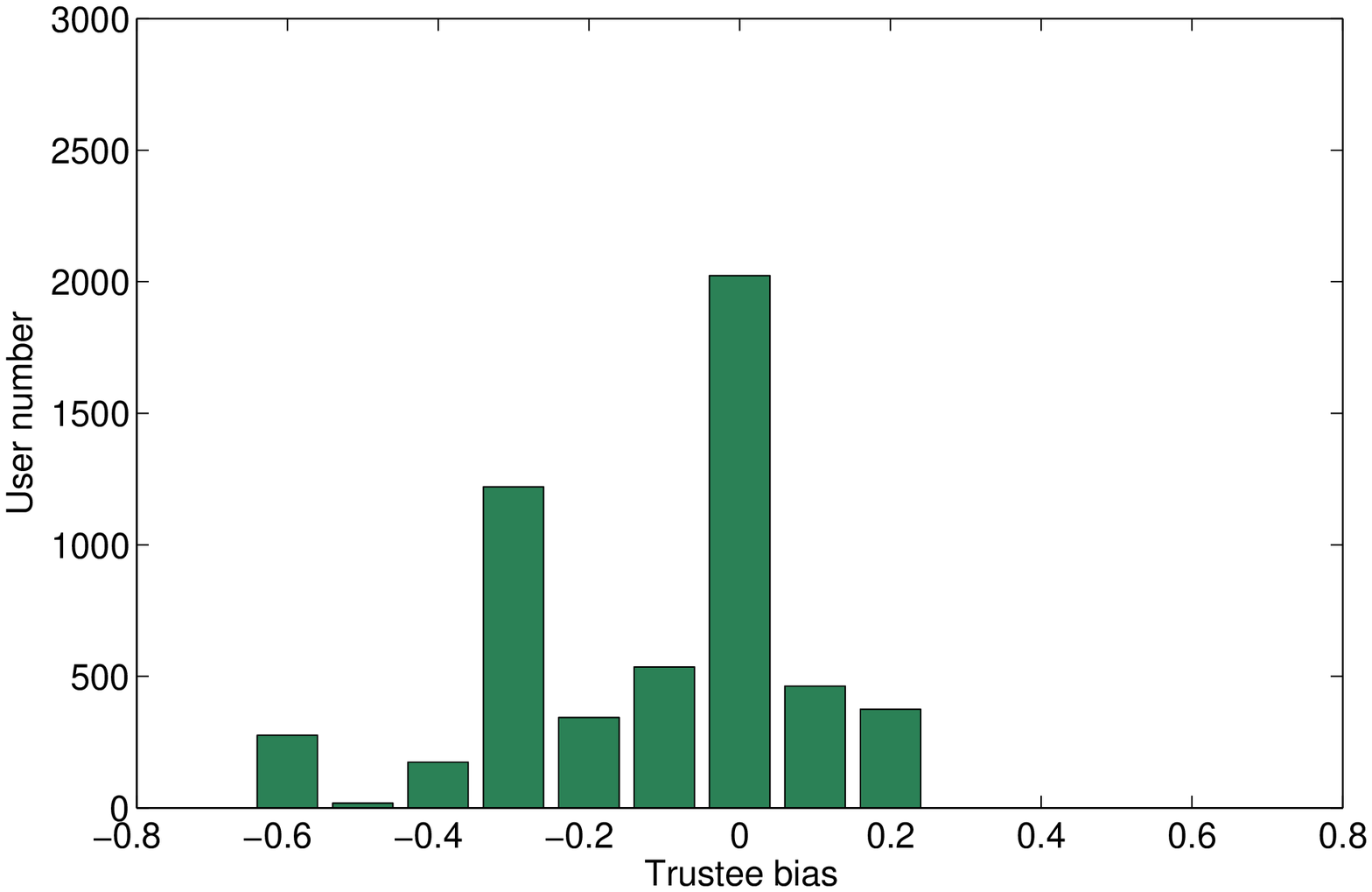}}
    \hspace{0.1in}
  \subfigure[Trustor bias distribution on PGP]{
  \label{F:trustorbias:pgp}\centering
    \includegraphics[width=1.5in]{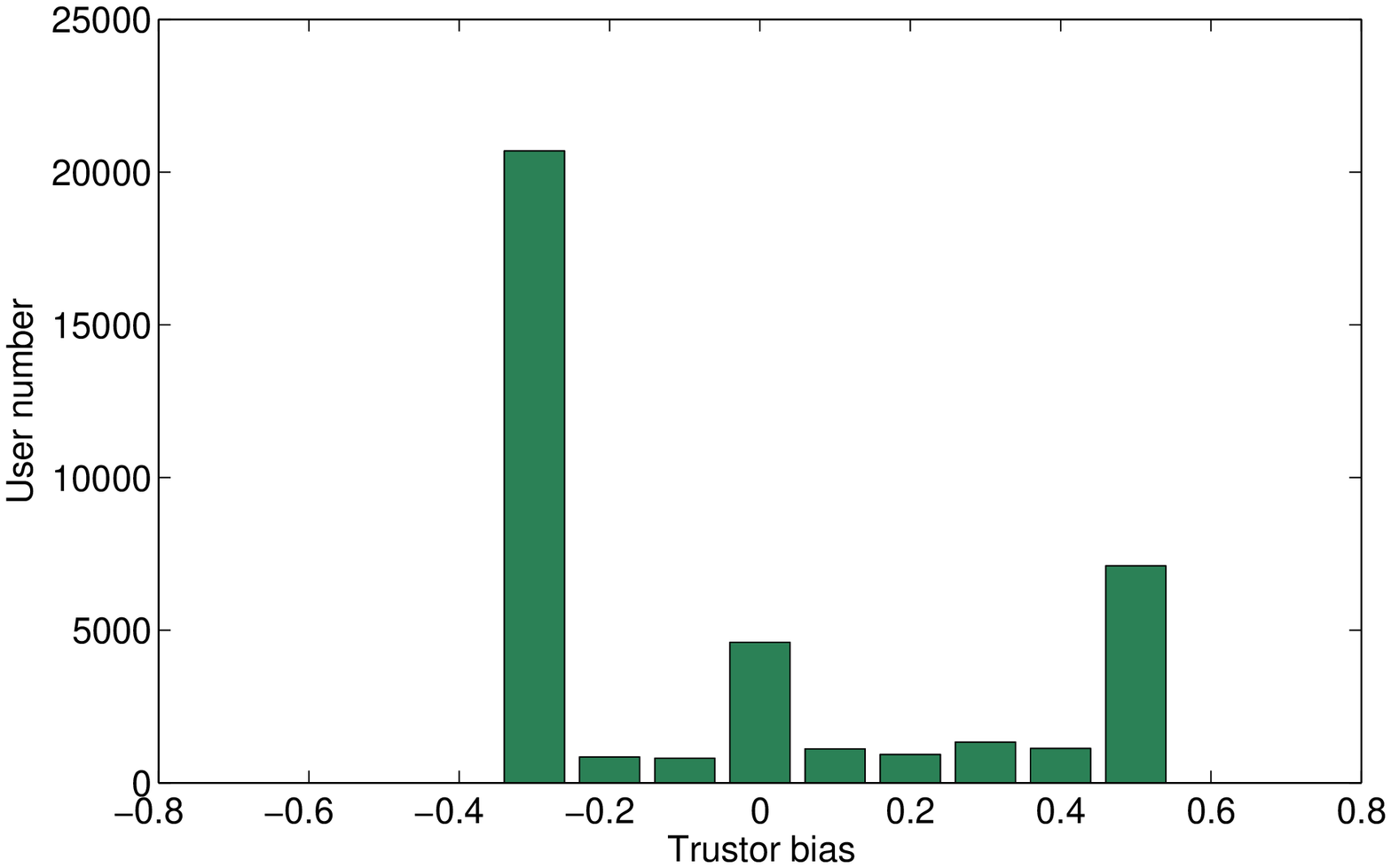}}
    \hspace{0.1in}
  \subfigure[Trustee bias distribution on PGP]{
  \label{F:trusteebias:pgp}\centering
    \includegraphics[width=1.5in]{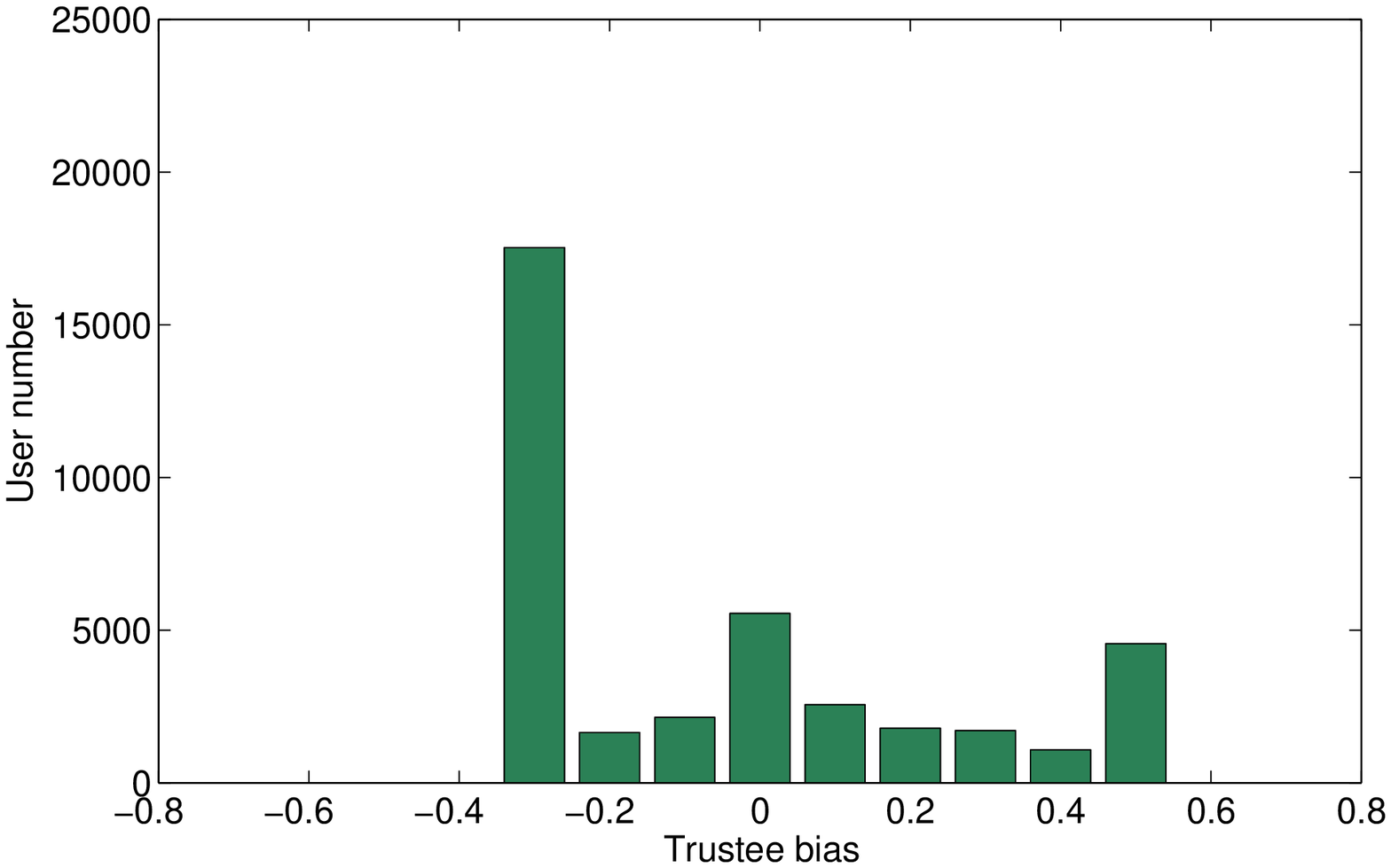}}
  \caption{The distributions of trustor bias and trustee bias.}
\label{F:biasdistribution}\centering
\end{figure}

Fig.~\ref{F:biasdistribution} presents the distributions of trustor bias and trustee bias. As we can see, many users in adovogato perform averagely on trusting others and being trusted by others. On the other hand, a considerable part of PGP users are cautiously trusted by others, and even more users tend to rate others strictly. The global bias is 0.6679 and 0.3842 for advogato and PGP, respectively. This also confirms that the security-related PGP network is a more conservative environment than the developer-based advogato network.

\subsection{Effectiveness Results}

\begin{figure}[!t]
  \centering
  \subfigure[{\em advogato} data set]{
  \label{F:effectiveness:advogato}\centering
    \includegraphics[width=2.5in]{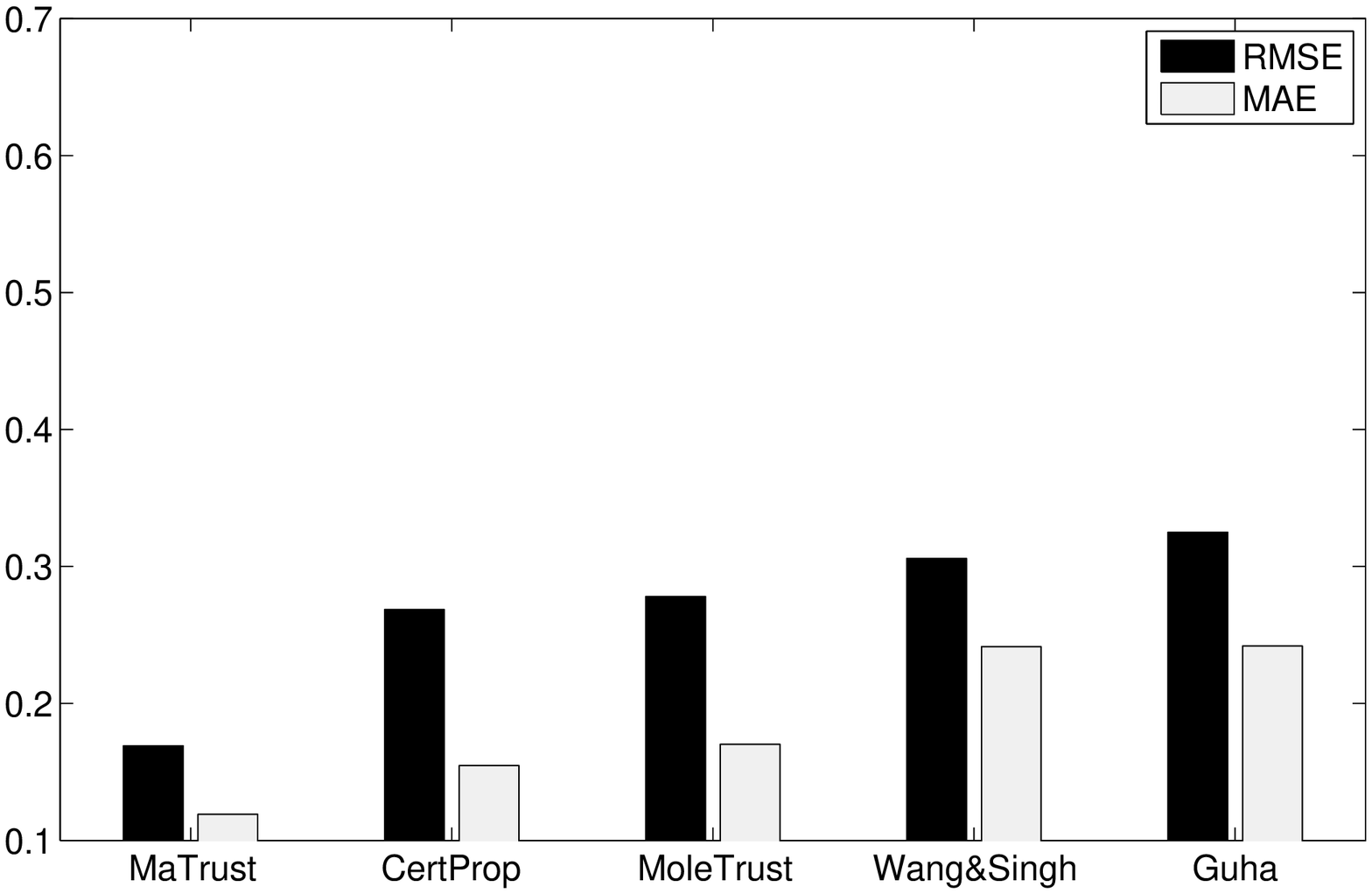}}
    \hspace{0.1in}
  \subfigure[{\em PGP} data set]{
  \label{F:effectiveness:pgp}\centering
    \includegraphics[width=2.5in]{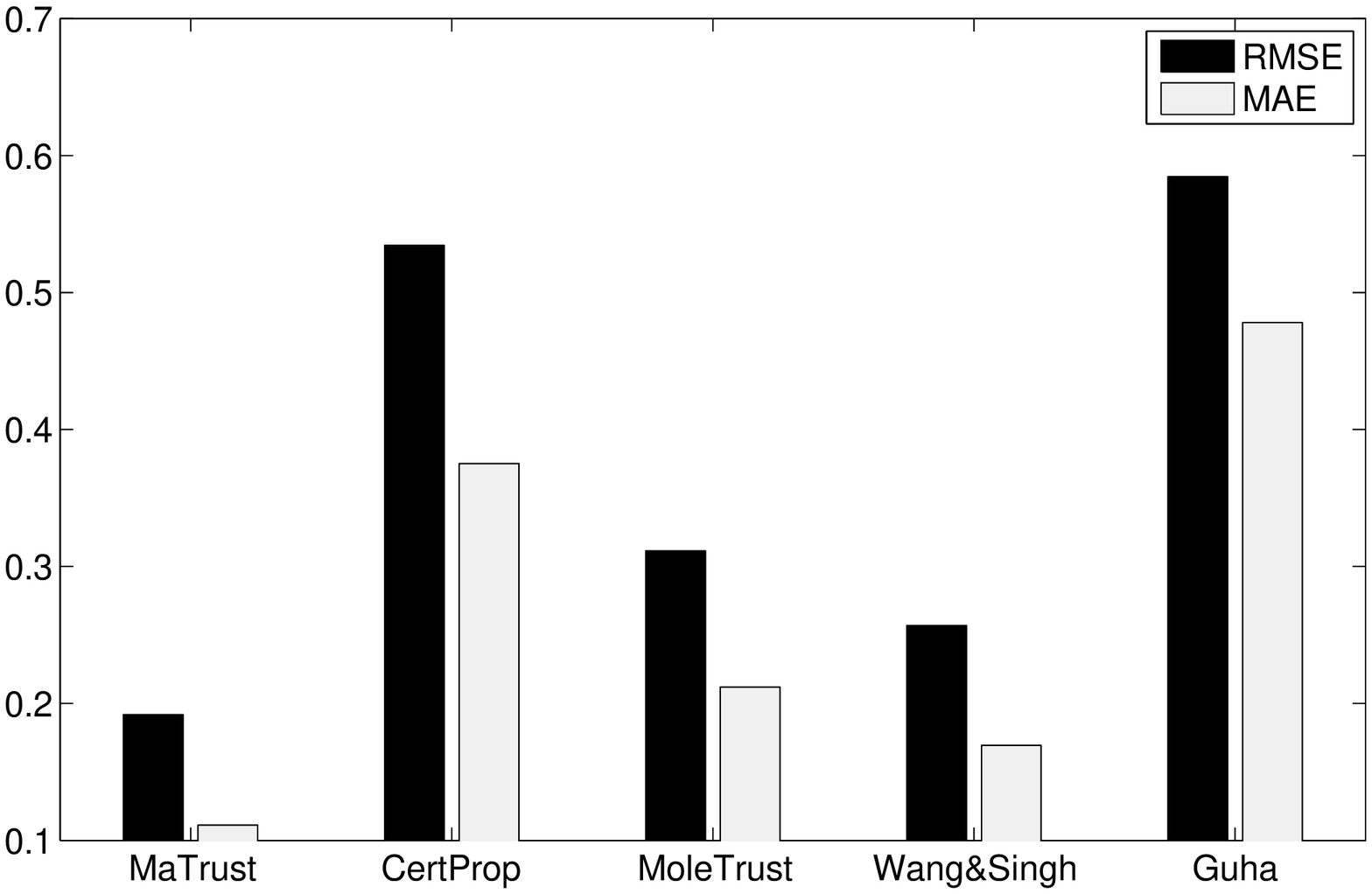}}
  \caption{Comparisons with subjective trust inference models. The proposed MaTrust significantly outperforms all the other models wrt both RMSE and MAE on both data sets.}
\label{F:effectiveness}\centering
\end{figure}

We use both {\em advogato} (i.e., {\em advogato-6}) and {\em PGP} for effectiveness evaluations. For both data sets, we hide a randomly selected sample of 500 observed trustor-trustee pairs as the test set, and apply the proposed MaTrust as well as other compared methods on the remaining data set to infer the trustworthiness scores for those hidden pairs. To evaluate and compare the accuracy, we report both the root mean squared error (RMSE) and the mean absolute error (MAE) between the estimated and the true trustworthiness scores. Both RMSE and MAE are measured on the 500 hidden pairs in the test set. We set $\lambda = 1.0$, $r = 10$, $m_1 = 10$, $m_2 = 100$, and $\xi_1 = \xi_2 = 10^{-6}$ in our experiments unless
otherwise specified.

{\em (A) Comparisons with Existing Subjective Trust Inference Methods}.  We first compare the effectiveness of MaTrust with several benchmark trust propagation models, including \emph{CertProp}~\cite{hang2009operators}, \emph{MoleTrust}~\cite{massa2005controversial}, \emph{Wang\&Singh}~\cite{wang2006trust,wang2007formal}, and \emph{Guha}~\cite{guha2004propagation}. For all these subjective methods, the goal is to infer a pair-wise trustworthiness score (i.e., to what extent the user $u$ trusts another user $v$).

The result is shown in Fig.~\ref{F:effectiveness}. We can see that the proposed MaTrust significantly outperforms all the other trust inference models wrt both RMSE and MAE on both data sets. For example, on {\em advogato} data set, our MaTrust improves the best existing method (CertProp) by 37.1\% in RMSE and by 23.0\% in MAE. As for {\em PGP} data set, the proposed MaTrust improves the best existing method (Wang\&Singh) by 25.3\% in RMSE and by 34.3\% in MAE. The results suggest that multi-aspect of trust indeed plays a very important role in the inference process.


{\em (B) Comparisons with Existing Objective Trust Inference Methods}.  Although our MaTrust is a subjective trust inference metric, as a side product, it can also be used to infer an objective trustworthiness score for each trustee. To this end, we set $r=1$ in MaTrust algorithm, and aggregate the resulting trustee matrix/vector $\mat {G}_0$ with the bias (the global bias $\mu$ and the trustee bias $\textbf{y}$). We compare the result with a widely-cited objective trust inference model {\em EigenTrust}~\cite{kamvar2003eigentrust} in Table~\ref{T:eigentrust}. As we can see, MaTrust outperforms EigenTrust in terms of both RMSE and MAE on both data sets. For example, on {\em advogato} data set, MaTrust is 58.6\% and 68.9\% better than EigenTrust wrt RMSE and MAE, respectively.

\begin{table}[!t]
\caption{Comparisons with {\em EigenTrust}. MaTrust is better than EigenTrust wrt both RMSE and MAE on both data sets.}
\label{T:eigentrust}\centering
\begin{tabular}{c||c|c}
  \hline
  RMSE/MAE & advogato & PGP \\ \hline\hline
  EigenTrust & 0.700 / 0.653 & 0.519 / 0.371 \\
  MaTrust & 0.290 / 0.203 & 0.349 / 0.280 \\\hline
\end{tabular}
\end{table}

\begin{table}[!t]
\caption{The importance of trust bias. Trust bias significantly improves trust inference accuracy.}
\label{T:bias}\centering
\begin{tabular}{c||c|c}
  \hline
  RMSE/MAE & advogato & PGP \\ \hline\hline
  MaTrust without trust bias & 0.228 / 0.164 & 0.244 / 0.135 \\
  MaTrust & 0.169 / 0.119 & 0.192 / 0.111 \\\hline
\end{tabular}
\end{table}

\begin{table}[!t]
\caption{Comparisons with {\em SVD} and {\em KBV}~\cite{koren2009matrix}. MaTrust outperforms both of them.}
\label{T:koren}\centering
\begin{tabular}{c||c|c}
  \hline
  RMSE/MAE & advogato & PGP \\ \hline\hline
  SVD & 0.629 / 0.579 & 0.447 / 0.306 \\
  KBV & 0.179 / 0.125 & 0.217 / 0.133 \\
  MaTrust & 0.169 / 0.119 & 0.192 / 0.111 \\\hline
\end{tabular}
\end{table}

{\em (C) Trust Bias Evaluations}. We next show the importance of trust bias by comparing MaTrust with the results when trust bias is not incorporated. The result is shown in Table~\ref{T:bias}. As we can see, MaTrust performs much better when trust bias is incorporated. For example, on {\em PGP} data set, trust bias helps MaTrust to obtain 21.3\% and 17.8\% improvements in RMSE and MAE, respectively. This result confirms that trust bias also plays an important role in trust inference.

{\em (D) Comparisons with Existing Matrix Factorization Methods}. We also compare MaTrust with two existing matrix factorization methods: {\em SVD} and the collaborative filtering algorithm~\cite{koren2009matrix} for recommender systems (referred to as {\em KBV}).

The result is shown in Table~\ref{T:koren}. As we can see from the table, MaTrust again outperforms both SVD and KBV on both data sets. {\em SVD} performs poorly as it treats all the unobserved trustor-trustee pairs as zero elements in the trust matrix $\mat T$. MaTrust also outperforms {\em KBV}. For example, MaTrust improves {\em KBV} by 11.5\% in RMSE and by 16.5\% in MAE on {\em PGP} data set. As mentioned before, {\em KBV} can be viewed as a special case of the proposed MaTrust if we fix all the coefficients as $1$.  This result confirms that by simultaneously learning the bias coefficients from the input trust matrix $\mat T$ (i.e., the relative weights for different types of bias), MaTrust leads to further performance improvement.

\begin{figure}[!t]
  \centering
  \subfigure[RMSE and MAE of MaTrust wrt $r$. We fix $r = 10$.]{
  \label{F:rank}\centering
    \includegraphics[width=2.7in]{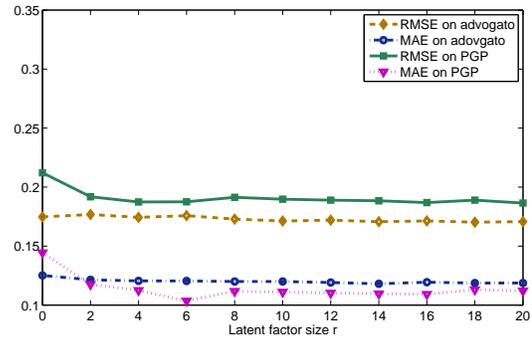}}
    \hspace{0.1in}
  \subfigure[RMSE and MAE of MaTrust wrt $\lambda$. We fix $\lambda = 1.0$.]{
  \label{F:lamda}\centering
    \includegraphics[width=2.7in]{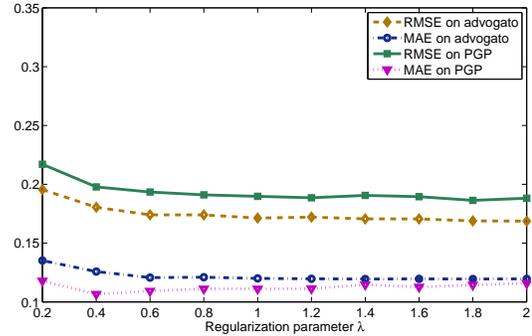}}
  \caption{The sensitivity evaluations. MaTrust is robust wrt both parameters.}
\label{F:parameters}\centering
\end{figure}

{\em (E) Sensitivity Evaluations}. Finally, we conduct a parametric study for MaTrust. The first parameter is the latent factor size $r$. We can observe from Fig.~\ref{F:rank} that, in general, both RMSE and MAE stay stable wrt $r$ with a slight decreasing trend. For example, compared with the results of $r=2$, the RMSE and MAE decrease by 3.1\% and 4.3\% on average if we increase $r=20$.  The second parameter in MaTrust is the regularization coefficient $\lambda$. As we can see from Fig.~\ref{F:lamda}, both RMSE and MAE decrease when $\lambda$ increases up to $0.8$; and they stay stable after $\lambda > 0.8$. Based on these results, we conclude that MaTrust is robust wrt both parameters. For all the other results we report in the paper, we simply fix $r=10$ and $\lambda=1.0$.

\subsection{Efficiency Results}

For efficiency experiments, we report the average wall-clock time. All the experiments were run on a machine with two 2.4GHz Intel Cores and 4GB memory.

\begin{figure}[!t]
  \centering
  \subfigure[Wall-clock time on {\em advogato} data set]{
  \label{F:efficiency:advogato}\centering
    \includegraphics[width=2.7in]{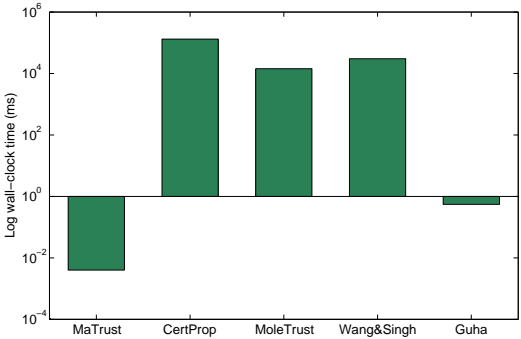}}
    \hspace{0.1in}
  \subfigure[Wall-clock time on {\em PGP} data set]{
  \label{F:efficiency:pgp}\centering
    \includegraphics[width=2.7in]{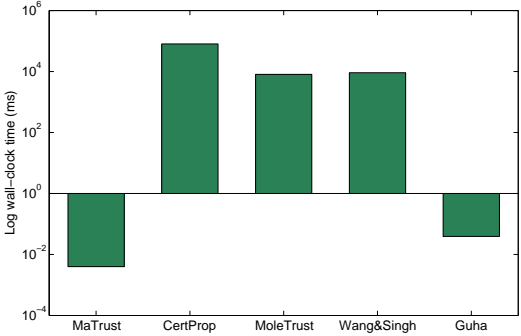}}
  \caption{Speed comparison. MaTrust is much faster than all the other methods.}
\label{F:efficiency}\centering
\end{figure}

{\em (A) Speed Comparison}.  We first compare the on-line response of MaTrust with \emph{CertProp}, \emph{MoleTrust}, \emph{Wang\&Singh}, and \emph{Guha}. Again, we use the {\em advogato-6} snapshot and {\em PGP} in this experiment, and the result is shown in Fig.~\ref{F:efficiency}. Notice that the y-axis is in the logarithmic scale. \hide{When applying CertProp, MoleTrust, and Wang\&Singh on PGP, we also shortened the propagation distance to make these models feasible.}

We can see from the figure that the proposed MaTrust is much faster than all the alternative methods on both data sets. For example, MaTrust is  2,000,000 - 3,500,000x faster than MoleTrust. This is because once we have inferred the trustor/truestee matrices as well as the coefficients for the bias (Step 1-12 in Algorithm~\ref{A:matrust}), it only takes {\em constant} time for MaTrust to output the trustworthiness score (Step 13 in  Algorithm~\ref{A:matrust}).
Among all the alternative methods, Guha is the most efficient. This is because its main workload can also be completed in advance. However, the pre-computation of Guha needs additional $O(n^2)$ space as the model fills nearly all the missing elements in the trust matrix, making it unsuitable for large data sets. In contrast, MaTrust only requires $O(|\mathcal{K}| + nr + r^2)$ space, which is usually much smaller than $n^2$.


\begin{figure}[!t]
  \centering
  \subfigure[Wall-clock time vs. $n$ on advogato]{
  \label{F:scalability:advogato}\centering
    \includegraphics[width=1.5in]{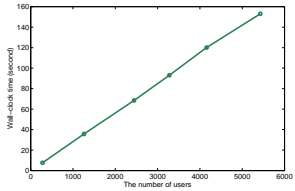}}
    \hspace{0.1in}
  \subfigure[Wall-clock time vs. $|{\cal{K}}|$ on advogato]{
  \label{F:scalability:advogato}\centering
    \includegraphics[width=1.5in]{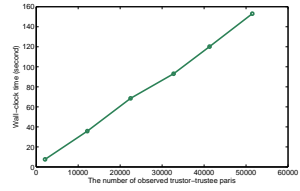}}
    \hspace{0.1in}
  \subfigure[Wall-clock time vs. $n$ on PGP]{
  \label{F:scalability:pgp}\centering
    \includegraphics[width=1.5in]{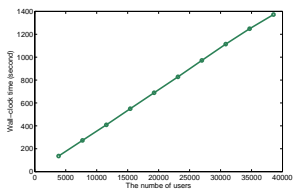}}
    \hspace{0.1in}
  \subfigure[Wall-clock time vs. $|{\cal{K}}|$ on PGP]{
  \label{F:scalability:pgp}\centering
    \includegraphics[width=1.5in]{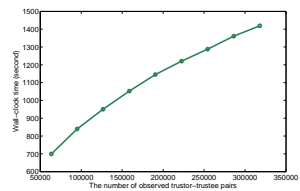}}
  \caption{Scalability of the proposed MaTrust. MaTrust scales linearly wrt the data size ($n$ and $|{\cal{K}}|$).}
\label{F:scalability}\centering
\end{figure}

{\em (B) Scalability}. Finally, we present the scalability result of MaTrust by reporting the wall-clock time of the pre-computational stage (i.e., Step 1-12 in Algorithm~\ref{A:matrust}). For {\em advogato} data set, we directly report the results on all the six snapshots (i.e., {\em advogato-1}, ..., {\em advogato-6}). For {\em PGP}, we use its subsets to study the scalability. The result is shown in Fig.~\ref{F:scalability}, which is consistent with the complexity analysis in Section~\ref{sec:algorithmanalysis}. As we can see from the figure, MaTrust scales linearly wrt to both $n$ and $|\mathcal{K}|$, indicating that it is suitable for large-scale applications.


\section{Conclusion}

In this paper, we have proposed an effective multi-aspect trust inference model (MaTrust). The key idea of MaTrust is to characterize several aspects/factors for each trustor and trustee based on the existing trust relationships. The proposed MaTrust can naturally incorporate the prior knowledge such as trust bias by expressing it as specified factors. In addition, MaTrust scales linearly wrt the input data size (e.g., the number of users, the number of observed trustor-trustee pairs, etc). Our experimental evaluations on real data sets show that trust bias can truly improve the inference accuracy, and that MaTrust significantly outperforms existing benchmark trust inference models in both effectiveness and efficiency.  Future work includes investigating the capability of MaTrust to address the distrust as well as the trust dynamics.



\bibliographystyle{abbrv}
\bibliography{factorization}

\begin{thebibliography}{10}

\bibitem{bell2007modeling}
R.~Bell, Y.~Koren, and C.~Volinsky.
\newblock Modeling relationships at multiple scales to improve accuracy of
  large recommender systems.
\newblock In {\em KDD}, pages 95--104. ACM, 2007.

\bibitem{buchanan2005damped}
A.~Buchanan and A.~Fitzgibbon.
\newblock Damped newton algorithms for matrix factorization with missing data.
\newblock In {\em CVPR}, volume~2, pages 316--322, 2005.

\bibitem{burnett2010bootstrapping}
C.~Burnett, T.~Norman, and K.~Sycara.
\newblock Bootstrapping trust evaluations through stereotypes.
\newblock In {\em AAMAS}, pages 241--248, 2010.

\bibitem{chiang2011exploiting}
K.~Chiang, N.~Natarajan, A.~Tewari, and I.~Dhillon.
\newblock Exploiting longer cycles for link prediction in signed networks.
\newblock In {\em CIKM}, pages 1157--1162, 2011.

\bibitem{gefen2002reflections}
D.~Gefen.
\newblock Reflections on the dimensions of trust and trustworthiness among
  online consumers.
\newblock {\em ACM SIGMIS Database}, 33(3):38--53, 2002.

\bibitem{golbeck2009trust}
J.~Golbeck.
\newblock Trust and nuanced profile similarity in online social networks.
\newblock {\em ACM Transactions on the Web}, 3(4):12, 2009.

\bibitem{guha2004propagation}
R.~Guha, R.~Kumar, P.~Raghavan, and A.~Tomkins.
\newblock {Propagation of trust and distrust}.
\newblock In {\em WWW}, pages 403--412. ACM, 2004.

\bibitem{hang2009operators}
C.-W. Hang, Y.~Wang, and M.~P. Singh.
\newblock {Operators for propagating trust and their evaluation in social
  networks}.
\newblock In {\em AAMAS}, pages 1025--1032, 2009.

\bibitem{hsieh2012low}
C.~Hsieh, K.~Chiang, and I.~Dhillon.
\newblock Low rank modeling of signed networks.
\newblock In {\em KDD}, 2012.

\bibitem{josang2002beta}
A.~J{\o}sang and R.~Ismail.
\newblock {The Beta reputation system}.
\newblock In {\em Proc. of the 15th Bled Electronic Commerce Conference},
  volume 160, Bled, Slovenia, June 2002.

\bibitem{kamvar2003eigentrust}
S.~D. Kamvar, M.~T. Schlosser, and H.~Garcia-Molina.
\newblock {The Eigentrust algorithm for reputation management in p2p networks}.
\newblock In {\em WWW}, pages 640--651. ACM, 2003.

\bibitem{koren2009matrix}
Y.~Koren, R.~Bell, and C.~Volinsky.
\newblock Matrix factorization techniques for recommender systems.
\newblock {\em Computer}, 42(8):30--37, 2009.

\bibitem{kuter2007sunny}
U.~Kuter and J.~Golbeck.
\newblock {Sunny: A new algorithm for trust inference in social networks using
  probabilistic confidence models}.
\newblock In {\em AAAI}, pages 1377--1382, 2007.

\bibitem{leskovec2010predicting}
J.~Leskovec, D.~Huttenlocher, and J.~Kleinberg.
\newblock Predicting positive and negative links in online social networks.
\newblock In {\em WWW}, pages 641--650. ACM, 2010.

\bibitem{leskovec2005graphs}
J.~Leskovec, J.~Kleinberg, and C.~Faloutsos.
\newblock Graphs over time: densification laws, shrinking diameters and
  possible explanations.
\newblock In {\em KDD}, pages 177--187. ACM, 2005.

\bibitem{liu2010optimal}
G.~Liu, Y.~Wang, and M.~Orgun.
\newblock Optimal social trust path selection in complex social networks.
\newblock In {\em AAAI}, pages 1391--1398, 2010.

\bibitem{liu2011trust}
G.~Liu, Y.~Wang, and M.~Orgun.
\newblock Trust transitivity in complex social networks.
\newblock In {\em AAAI}, pages 1222--1229, 2011.

\bibitem{liu2009stereotrust}
X.~Liu, A.~Datta, K.~Rzadca, and E.~Lim.
\newblock Stereotrust: a group based personalized trust model.
\newblock In {\em CIKM}, pages 7--16. ACM, 2009.

\bibitem{ma2009learning}
H.~Ma, M.~Lyu, and I.~King.
\newblock Learning to recommend with trust and distrust relationships.
\newblock In {\em RecSys}, pages 189--196. ACM, 2009.

\bibitem{massa2005controversial}
P.~Massa and P.~Avesani.
\newblock {Controversial users demand local trust metrics: An experimental
  study on epinions. com community}.
\newblock In {\em AAAI}, pages 121--126, 2005.

\bibitem{mishra2011finding}
A.~Mishra and A.~Bhattacharya.
\newblock Finding the bias and prestige of nodes in networks based on trust
  scores.
\newblock In {\em WWW}, pages 567--576. ACM, 2011.

\bibitem{nguyen2009trust}
V.~Nguyen, E.~Lim, J.~Jiang, and A.~Sun.
\newblock To trust or not to trust? predicting online trusts using trust
  antecedent framework.
\newblock In {\em ICDM}, pages 896--901. IEEE, 2009.

\bibitem{nordheimer2010trustworthiness}
K.~Nordheimer, T.~Schulze, and D.~Veit.
\newblock Trustworthiness in networks: A simulation approach for approximating
  local trust and distrust values.
\newblock In {\em IFIPTM}, volume 321, pages 157--171. Springer-Verlag, 2010.

\bibitem{resnick2002trust}
R.~Paul and Z.~Richard.
\newblock {Trust among strangers in Internet transactions: Empirical analysis
  of eBay's reputation system}.
\newblock In {\em The Economics of the Internet and E-Commerce}, volume~11 of
  {\em Advances in Applied Microeconomics: A Research Annual}, pages 127--157.
  Elsevier, 2002.

\bibitem{richardson2003trust}
M.~Richardson, R.~Agrawal, and P.~Domingos.
\newblock {Trust management for the Semantic Web}.
\newblock In {\em ISWC}, pages 351--368. Springer, 2003.

\bibitem{sabater2002reputation}
J.~Sabater and C.~Sierra.
\newblock {Reputation and social network analysis in multi-agent systems}.
\newblock In {\em AAMAS}, pages 475--482. ACM, 2002.

\bibitem{sirdeshmukh2002consumer}
D.~Sirdeshmukh, J.~Singh, and B.~Sabol.
\newblock Consumer trust, value, and loyalty in relational exchanges.
\newblock {\em The Journal of Marketing}, pages 15--37, 2002.

\bibitem{tang2012mtrust}
J.~Tang, H.~Gao, and H.~Liu.
\newblock m{T}rust: discerning multi-faceted trust in a connected world.
\newblock In {\em WSDM}, pages 93--102. ACM, 2012.

\bibitem{tversky1974judgment}
A.~Tversky and D.~Kahneman.
\newblock Judgment under uncertainty: Heuristics and biases.
\newblock {\em science}, 185(4157):1124--1131, 1974.

\bibitem{wang2011multi}
G.~Wang and J.~Wu.
\newblock Multi-dimensional evidence-based trust management with multi-trusted
  paths.
\newblock {\em Future Generation Computer Systems}, 27(5):529--538, 2011.

\bibitem{wang2006trust}
Y.~Wang and M.~P. Singh.
\newblock {Trust representation and aggregation in a distributed agent system}.
\newblock In {\em AAAI}, pages 1425--1430, 2006.

\bibitem{wang2007formal}
Y.~Wang and M.~P. Singh.
\newblock {Formal trust model for multiagent systems}.
\newblock In {\em IJCAI}, pages 1551--1556, 2007.

\bibitem{watts1998collective}
D.~Watts and S.~Strogatz.
\newblock {Collective dynamics of 'small-world' networks}.
\newblock {\em Nature}, 393(6684):440--442, 1998.

\bibitem{xiong2004peertrust}
L.~Xiong and L.~Liu.
\newblock {Peertrust: Supporting reputation-based trust for peer-to-peer
  electronic communities}.
\newblock {\em IEEE Transactions on Knowledge and Data Engineering},
  16(7):843--857, 2004.

\bibitem{yao2012subgraph}
Y.~Yao, H.~Tong, F.~Xu, and J.~Lu.
\newblock Subgraph extraction for trust inference in social networks.
\newblock In {\em ASONAM (to appear)}, 2012.

\bibitem{ziegler2005propagation}
C.~Ziegler and G.~Lausen.
\newblock Propagation models for trust and distrust in social networks.
\newblock {\em Information Systems Frontiers}, 7(4):337--358, 2005.

\end{thebibliography}

\appendix
\section{Detailed Algorithm~1}
Here, we present the complete algorithm to update the trustor/trustee matrices when the bias coefficients are fixed (i.e., Algorithm~\ref{A:skeleton} for Eq.~\eqref{E:optsimple}). As mentioned above, we apply the alternating strategy by alternatively fixing one of the two matrices and optimizing the other. For simplicity, let us consider how to update $\textbf{F}_0$ when $\textbf{G}_0$ is fixed. Updating $\textbf{G}_0$ when $\textbf{F}_0$ is fixed can be done in a similar way. By fixing $\textbf{G}_0$, Eq.~\eqref{E:optsimple} can be further simplified as follows: \begin{equation}\label{E:optf}
\min_{\textbf{F}_0} \sum_{(i,j)\in \mathcal{K}} (\textbf{P}(i,j) - \textbf{F}_0(i,:)\textbf{G}_0(j,:)')^2 + \lambda ||\textbf{F}_0||_{fro}^2
\end{equation}

In fact, the above optimization problem in Eq.~\eqref{E:optf} now becomes convex wrt $\textbf{F}_0$. It can be further decoupled into many independent sub-problems, each of which only involves a single row in $\mat F_0$: \begin{equation}\label{E:optfline}
\min_{\textbf{F}_0(i,:)} \sum_{j,(i,j)\in \mathcal{K}} (\textbf{P}(i,j) - \textbf{F}_0(i,:)\textbf{G}_0(j,:)')^2 + \lambda ||\textbf{F}_0(i,:)||^2
\end{equation}

The optimization problem in Eq.~\eqref{E:optfline} can now be solved by the standard ridge regression wrt the corresponding row $\textbf{F}_0(i,:)$.

\begin{algorithm}[h]
\caption{alternateUpdate($\textbf{P}, \textbf{F}_0, \textbf{G}_0$).}\label{A:alternating}
\begin{algorithmic}[1]
  \REQUIRE {The $n \times n$ matrix $\textbf{P}$, the $n \times r$ matrix $\textbf{F}_0$, and the fixed $n \times r$ matrix $\textbf{G}_0$}
  \ENSURE {The updated matrix $\textbf{F}_1$ of $\textbf{F}_0$}
    \STATE $\textbf{F}_1$ $\leftarrow$ $\textbf{F}_0$;
    \FOR {i = 1 : n}
      \STATE $\textbf{a}$ $\leftarrow$ the vector of column indices of existing elements in $\textbf{P}(i,j)~(j=1,2,...,n)$;
      \STATE column vector $\textbf{d}$ $\leftarrow$ $\textbf{0}_{|\textbf{a}| \times 1}$;
      \STATE matrix $\textbf{G}_1$ $\leftarrow$ $\textbf{0}_{|\textbf{a}| \times r}$;
      \FOR {j = 1: $|\textbf{a}|$}
        \STATE $\textbf{d}(j)$ $\leftarrow$ $\textbf{P}(i,\textbf{a}(j))$; 
        \STATE $\textbf{G}_1(j,:)$ $\leftarrow$ $\textbf{G}_0(\textbf{a}(j), :)$; 
      \ENDFOR
      \STATE $\textbf{F}_1(i,:)$ $\leftarrow$ $(\textbf{G}'_1 \textbf{G}_1 + \lambda \cdot \textbf{I}_{r \times r})^{-1} \textbf{G}'_1 \textbf{d}$;
    \ENDFOR
  \RETURN $\textbf{F}_1$;
\end{algorithmic}
\end{algorithm}

Algorithm~\ref{A:alternating} presents the overall solution for updating the trustor matrix $\mat{F}_0$. Based on Algorithm~\ref{A:alternating}, we present Algorithm~\ref{A:factorization} to alternatively update the trustor and trustee matrices $\textbf{F}_0$ and $\textbf{G}_0$. The algorithm first generates two $n \times r$ matrices for $\textbf{F}_0$ and $\textbf{G}_0$ where each element is initialized as $1/r$. At each iteration, the algorithm then alternatively calls Algorithm~\ref{A:alternating} to update the two matrices. The iteration ends when the stopping criteria are met, i.e., either the $L_2$ norm between successive estimates of both $\textbf{F}_0$ and $\textbf{G}_0$ is below our threshold $\xi_2$ or the maximum iteration step $m_2$ is reached.

\begin{algorithm}[t]
\caption{updateMatrix(\textbf{P}, $r$).}\label{A:factorization}
\begin{algorithmic}[1]
  \REQUIRE {The $n \times n$ matrix $\textbf{P}$, and the latent factor size $r$}
  \ENSURE {The $n \times r$ trustor matrix $\textbf{F}_0$, and the $n \times r$ trustee matrix $\textbf{G}_0$}
  \STATE generate the $n \times r$ matrices $\textbf{F}_0$ and $\textbf{G}_0$ randomly;
  \WHILE {not convergent}
    \STATE $\textbf{F}_0$ $\leftarrow$ alternateUpdate($\textbf{P}$, $\textbf{F}_0$, $\textbf{G}_0$);
    \STATE $\textbf{G}_0$ $\leftarrow$ alternateUpdate($\textbf{P}'$, $\textbf{G}_0$, $\textbf{F}_0$);
  \ENDWHILE
  \RETURN [$\textbf{F}_0$, $\textbf{G}_0$];
\end{algorithmic}
\end{algorithm}

\section{Proofs for Lemmas}
Next, we present the proofs for the lemmas in Section~\ref{sec:algorithmanalysis}.

\noindent \textbf{Proof Sketch for Lemma~\ref{L:effectiveness}:} (P1) First, Eq.~\eqref{E:optfline} is convex and therefore Step 10 in Algorithm~\ref{A:alternating} finds the global optima for updating a single row in the matrix $\mat{F}_0$. Notice that the optimization problem in Eq.~\eqref{E:optfline} is equivalent to that in Eq.~\eqref{E:optf}, and thus we have proved that Algorithm~\ref{A:alternating} finds the global optimal solution for the optimization problem in Eq.~\eqref{E:optf}.

(P2) Next, based on (P1) and the alternating procedure in Algorithm~\ref{A:factorization}, we have that Algorithm~\ref{A:factorization} finds a local minima for the optimization problem in Eq.~\eqref{E:optsimple}.

(P3) Finally, based on (P2) and the alternating procedure in Algorithm~\ref{A:matrust}, Lemma~\ref{L:effectiveness} holds. \QED

\noindent \textbf{Proof of Lemma~\ref{L:efficiency}:} (P1) In Algorithm~\ref{A:alternating}, the time cost for Step~1 is $O(n r)$. Let $a_i$ denote the number of elements in $\textbf{a}$ of the $i^{th}$ iteration. The time cost for Step~3-5 is then $O(a_i r)$ since we store $\textbf{P}$ in sparse format. We need another $O(a_i r)$ time in the inner iteration (Step~6-9). The time cost of Step~10 is $O(a_i r^2 + r^2 + r^3 + a_i r^2 + a_i r + r) = O(r^3+a_i r^2)$. Therefore, the total time cost for the algorithm is $O(n r) + O(\sum_i (r^3 + a_i r^2)) = O(nr^3 + |\mathcal{K}| r^2)$ where $\sum_i a_i = |\mathcal{K}|$.

(P2) In Algorithm~\ref{A:factorization}, the time cost for Step~1 is $O(n r)$. As indicated by (P1), we need $O(n r^3 + |\mathcal{K}| r^2)$ time for both Step~3 and Step~4. The total time cost is $O(n r^3 m_2 + |\mathcal{K}| r^2 m_2)$.

(P3) In Algorithm~\ref{A:matrust}, the time cost for Step~1 is $O(|\mathcal{K}|)$ as we store $\textbf{T}$ in sparse format. Step~2 needs $O(1)$ time. We need $O(|\mathcal{K}|)$ time for Step~4-6. As indicated by (P2), we need $O(n r^3 m_2 + |\mathcal{K}| r^2 m_2)$ time for Step~7. We need $O(|\mathcal{K}| r)$ time for Step~8-10. As for updating the coefficients, we need $O(|\mathcal{K}| c^2 + c^3)$ time where $c$ is the number of specified factors, which is $3$ in our case. Therefore, the time cost for Step~11 is $O(|\mathcal{K}|)$. The total time cost is $O(n r^3 m_1 m_2 + |\mathcal{K}| r^2 m_1 m_2)$, which completes the proof. \QED

\noindent \textbf{Proof of Lemma~\ref{L:efficiency2}:} (P1) In Algorithm~\ref{A:alternating}, we need $O(nr)$ space for Step~1 and $O(1)$ space for Step~2. We need another $O(nr)$ space for Step 3-5. For Step 6-9 we only need $O(1)$ space. We need $O(nr + r^2)$ space for Step~10. Among the different iterations of the algorithm, we can re-use the space from the previous iteration. Finally, the overall space cost is $O(|\mathcal{K}| + nr + r^2)$.

(P2) In Algorithm~\ref{A:factorization}, we need $O(nr)$ space for Step~1. Step~3 and Step~4 need $O(|\mathcal{K}| + nr + r^2)$ space. The space for each iteration can be re-used. The total space cost is $O(|\mathcal{K}| + nr + r^2)$.

(P3) In Algorithm~\ref{A:matrust}, we need $O(|\mathcal{K}|)$ space for the input since we store $\textbf{T}$ in sparse format. We need $O(n)$ space for Step~1 and $O(1)$ space for the Step~2. We need another $O(|\mathcal{K}|)$ space for Step~4-6. By (P2), Step~7 needs $O(|\mathcal{K}| + nr + r^2)$ space. Step 8-10 can re-use the space from Step 4-6. Step~11 needs $O(|\mathcal{K}|)$ space. For each iteration, the space can be re-used. The total space cost of Algorithm~\ref{A:matrust} is $O(|\mathcal{K}| + nr + r^2)$, which completes the proof. \QED


\balance
\end{document}